\documentclass[a4paper]{article}
\usepackage[width=16cm,height=26cm]{geometry}

\usepackage{helvet}
\usepackage[T1]{fontenc}	
\usepackage{amsmath}
\usepackage{amssymb}
\usepackage{mathrsfs}
\usepackage{graphics}
\usepackage[authoryear,square]{natbib}
\usepackage[usenames,dvipsnames]{color}
\usepackage{fancyhdr} 
\usepackage{subfigure} 
\usepackage[dvips]{graphicx} 
\usepackage{marvosym}
\usepackage{tikz} 
\usetikzlibrary{shapes,arrows}
\usepackage{framed}
\usepackage{titlesec}
\usepackage[simplified]{pgf-umlcd}
\usepackage{algorithm} 
\usepackage{algorithmic} 

\usepackage[normalem]{ulem}

\usepackage{listings}

\usepackage[outdir=./]{epstopdf}
\usepackage[section]{placeins}

\usepackage[nodisplayskipstretch]{setspace}

\definecolor{DarkGreen}{rgb}{0.0,0.4,0.0} 
\definecolor{highlight}{RGB}{255,251,204} 

\lstdefinestyle{Style1}{ 
	language=Perl, 
	backgroundcolor=\color{highlight}, 
	basicstyle=\footnotesize\ttfamily, 
	breakatwhitespace=false, 
	breaklines=true, 
	captionpos=b, 
	commentstyle=\usefont{T1}{pcr}{m}{sl}\color{DarkGreen}, 
	deletekeywords={}, 
	firstnumber=1, 
	frame=single, 
	frameround=tttt, 
	keywordstyle=\color{Blue}\bf, 
	morekeywords={}, 
	numbers=left, 
	numbersep=10pt, 
	numberstyle=\tiny\color{Gray}, 
	rulecolor=\color{black}, 
	showstringspaces=false, 
	showtabs=false, 
	stepnumber=5, 
	stringstyle=\color{Purple}, 
	tabsize=2, 
}

\newcommand{\norm}[1]{\left\lVert#1\right\rVert}

\newcommand{\PG}[1]{#1}

\title{Parallel Multiphysics Simulation for the Stabilized Optimal Transportation Meshfree (OTM) \PG{Method}}

\author{Sandeep Kumar$^1$, Pierre Gosselet$^4$, Dengpeng Huang$^2$, \\Christian Wei{\ss}enfels$^3$, Peter Wriggers$^1$
\\
\small
$^1$ Institute of Continuum Mechanics, Leibniz Universit{\"a}t Hannover, 30167 Hannover, Germany 
\\\small\texttt{ kumar@ikm.uni-hannover.de, wriggers@ikm.uni-hannover.de}\\
\small$^2$ Institute of Applied Dynamics, Friedrich-Alexander-Universit{\"a}t \\\small Erlangen-N{\"u}rnberg, 91058 Erlangen, Germany \\
\small\texttt{dengpeng.huang@fau.de}\\
\small$^3$  Institute of Materials Resource Management,\\\small Data-driven Computational Materials Science and Engineering,\\\small Universit{\"a}t Augsburg, 86159 Augsburg, Germany 
\\\small\texttt{christian.weissenfels@mrm.uni-augsburg.de}\\
\small$^4$ Université de Lille, CNRS, Centrale Lille, UMR 9013 -- LaMcube -- F-59000 Lille, France 
\\\small\texttt{pierre.gosselet@univ-lille.fr}
}

\date{10.1016/j.jocs.2022.101739}

\begin{document}
	\include{definitions}

\maketitle

\abstract{This paper presents a parallel \PG{implementation} for the Optimal Transportation Meshfree (OTM) method on large CPU clusters. Communications are handled with the Message Passing Interface (MPI). The Recursive Coordinate Bisection (RCB) algorithm is utilized for domain decomposition and for implementing dynamic load-balancing strategy. This work involves three new concepts to reduce the computational efforts: Dynamic halo regions, Efficient data management strategies for ease of addition and deletion of nodes and material points using advanced STL container, and nearest neighborhood communication for detection of neighbors and communication. Also, Linked Cell approach has been implemented to further reduce the computational efforts. Parallel performance analysis is investigated for challenging multiphysics applications like Taylor rod impact and serrated chip formation process. \PG{Adequate scalability of parallel implementation for these applications is reported.}
	
}

\section{Introduction}




Processes involving large deformations, such as Additive Manufacturing or cutting, present a challenge while modeling with standard approximation tools like the Finite Element Method. 
If the Lagrangian description is used, these large deformations can result in severe mesh distortions. In this case adaptive remeshing procedures and mapping of state variables from one configuration to another are required. Inefficient computations and accumulated numerical errors can result.  Alternatively, meshfree methods seem quite adapted to such simulations. For instance, the Smoothed Particle Hydrodynamics (SPH) has shown big potential. A more recent solution scheme is the Optimal Transportation Meshfree (OTM) method  \citep{2010_li_habbal_ortiz}. This method is motivated by the Optimal Transportation Theory  \citep{2013_villani} integrated with local maximum entropy (LME) meshfree interpolation  \citep{2006_arroyo_ortiz} and material point sampling method  \citep{1994_sulsky_chen_schreyer,wessels_et_al_2019,wessels_et_al_2018}. A detailed analysis on the  LME meshfree interpolation applied to metal forming process is provided in \citep{cueto_chinesta_2013}. The advantage of the OTM method is the similar transition of the FEM to a meshfree method \citep{2010_li_habbal_ortiz}.
\medskip 





Realistic simulations of many engineering applications problems require large-scale computations. Computation on HPC clusters needs efficient and scalable codes.   In order to utilize the full potential of the computing power of multi-core architectures, it is necessary to exploit both the intra and inter-node parallelism. Different parallel programming models are developed over the years. An overview can be found in \citep{prims_et_al_2019}. Currently, existing parallel programming models are based on distributed memory and shared memory platforms. Message Passing Interface (MPI) is most widely used standard paradigm on distributed memory platforms but it can also be applied to shared memory nodes. Other parallelization approaches exists, such as, OpenMP (for shared memory platforms), CUDA and OpenCL (for graphics processing units (GPUs)). Coupling our approach with OpenMP is possible and this will be the subject of future work. \medskip

Development of MPI \PG{based software} requires carefully thought strategies for data-partitioning and data communication. Data-partitioning refers to the process of dividing the problem domain into smaller subdomains. Subdomain geometry affects the scalability since it is associated with equivalent work load through load-balancing. In order to achieve high scalability, subdomains are expected to contain same amount of work (load balancing) while minimizing the need for communications. Several approaches exist in field of non-overlapping domain decomposition methods to solve the challenging mechanical heterogeneous problems on massively parallel architectures. A new parallel mesh generation method has been developed by   \citep{gharbi_et_al_2020} which leads to subdomains with shape well-suited for Schur based domain decomposition methods, such as FETI  \citep{farhat_roux_1991} and BDD \citep{mandel_1993} solvers. 
Another domain decomposition method, Orthogonal Recursive Bisection (ORB) algorithm has been implemented by   \citep{oger_et_al_2016, yang_et_al_2020} which has been shown to lead to scalable results at large processor numbers. Also, due to the distributed nature of the \PG{method}, duplication of the data is required for communication, resulting in the increased overall memory requirement. \medskip


 Generally when a parallel program is run on several processes simultaneously, there are data dependencies between the tasks. A process might need intermediate results in order to carry out its computations and this intermediate result could be located on adifferent process. Hence, bottlenecks occur which slow down the computation. The MPI library provides different communication primitives: point-to-point and collective communication. One way to maximize the performance of parallelization is to reduce the  overheads due to communication operations. Overhead is defined as \textit{the length of time that a processor is engaged in the transmission or reception of each message; during this time, the processor cannot perform other operations} \citep{culler_et_al_1993}. Collective communication operations, introduced in the latter versions of MPI, have been a key concept used in large scale parallel applications to minimize the communication overheads  \citep{barigou_gabriel_2017,barigou_et_al_2015}. Although they are widely used due to their increased productivity and performance, there are some limitations. Due to the dependencies on all the processes of a communicator, there exist scalability issues and conventional collectives support limited communication patterns, such as, broadcast and all-to-all, see   \citep{ghazimirsaeed_et_al_2020}. In order to address these issues, Neighborhood Collectives, introduced by the MPI 3.0, provide an alternative to the users to define arbitrary communication patterns. This can be used to implement nearest neighbor collective operations where each process interacts with only a small neighborhood of processes. Neighborhoods can be described either by Cartesian neighborhoods or by general communication graphs, for more details see  \citep{hoefler_et_al_2011} and  \citep{mpi_standard_2015}. \medskip

When performing computations within each subdomain, all the necessary information should be available in the same process. But, for the subdomain's boundary, some of the required information will be located on other processes. Hence, a communication pattern is required, the most commonly used is halo regions. At every computation step, the halo regions are exchanged with the neighboring processes so that every process can access to the necessary information. The goal of the halo regions is to locally replicate the domain residing in other processes. 
At every computation step, when processes communicate with their neighbors, performance of any process depends strongly  on the performance of its neighbors. This can result in delays \citep{kemp_2015}. Additionally, the probability for  delays can increase with the number of processes. \medskip


Large scale computations requires an adaptive code to run efficiently on distributed memory systems. Good data management and domain decomposition are critical parameters. Parallelizing OTM‌ shares many common issues with discrete
element methods \citep{visseq_et_al_2013}. Some of the cumbersome tasks during simulation using the OTM method in a parallel environment involves modifying, deleting and adding particles in a subdomain, adjusting the subdomain partitioning dynamically and performing migration of particles to maintain load balance during the simulations. These tasks require flexible and efficient data management scheme. In mesh-based methods, flexible and efficient data management schemes for parallel systems have been implemented for adaptive hp finite element method \citep{Laszloffy_et_al_2000}, and for simulation tool for geophysical mass flows  \citep{patra_et_al_2005}. In meshfree methods,   \citep{cao_et_al_2017} developed data management strategies for a MPI parallel implementation of the SPH method to simulate volcano plumes.   \citep{ferrari_et_al_2009} used a flexible way in linked lists using pointers so that particles can be deleted or added during the simulation. Similar approach for modifying pointer-based information has been adopted in the current work.\medskip


The following three reasons motivate the work presented in this paper. First, while parallelization approach to OTM method has been implemented by \citep{2014_li_stalzer_ortiz}, efficient implementation strategies have been presented by introducing communication for both nodal and material point halo regions for localized updates within every subdomain. Second, with the use of improved data structures for halo regions, flexibility have introduced to handle variable workloads (dynamic halo regions). This is helpful when new nodal or material point quantities are added into the communication. Finally, with the use of nearest neighborhood communication for neighbor detection and communication, the communication costs are reduced even though total halo particles increases with increase in number of subdomains.\medskip

In this work, computational strategies are proposed for parallel processing of OTM Method using MPI (Message Passing Interface) for scalability on large-scale computer clusters. Mechanisms are presented for efficient addition or removal of nodes and material points from their corresponding influence and support domains respectively, thereby reducing computational overheads. Hence, storage issues related to the fixed-size arrays are  eliminated. Dynamic halo region is implemented that can handle variable workloads. Both nodes and material points within every subdomain and their corresponding influence and support domains are managed by STL map which can ensure quick and flexible access and modifications. In order to ensure good static and dynamic load balance, Recursive Coordinate Bisection (RCB) algorithm, a Cartesian based  decomposition method is used for both static and dynamic decomposition (dynamic load balancing). The RCB decomposition method facilitates good scalability by ensuring minimum interfacial surface area between the sub-domains. Parallel decomposition of spatial domain is carried out in such a way that each subdomain is physically compact and the computations can be performed locally at each process. The flexibility of our data access methodology, data structures, dynamic halo regions  enables efficient parallel implementation of OTM method. To reduce global communications, nearest neighbor communication operations are implemented using MPI collectives \citep{mpi_standard_2015}.\medskip


The outline of this paper is as follows. Section  \ref{sec:otm_method} contains a brief description of the OTM Method. Section  \ref{sec:MPI_intro} discusses the data structures, as well as the parallel implementation for domain decomposition and communication. Results from applying the presented \PG{method} to the Taylor rod impact and the serrated chip formation process are presented in Section  \ref{sec:parallel_performance}.

\section{Optimal Transportation Meshfree Algorithm}\label{sec:otm_method}

The Optimal Transportation Meshfree (OTM) method is an Updated Lagrangian formulation which can be used for both solid and fluid flow simulations based on  \citep{2010_li_habbal_ortiz}. OTM method can be viewed as an evolution of finite element method because the spatial domain under investigation is discretized by two types of points. The material points are used as integration points, where quantities like stress, strain, density, etc., are determined. At the nodal points, the primary variables are computed by solving  discretized equations of motion. A search algorithm establishes the connectivity between nodes and material points during the computation: the nodes associated with a material point form its support domain, whose shape is in general arbitrary. The material point values are determined with the help of basis functions. In general, maximum entropy shape functions  \citep{2006_arroyo_ortiz} are used. Since OTM method have some shortcomings, the stabilized formulation due to \citep{2017_weissenfels_wriggers} is used.  
The whole algorithm is sketched in Algorithm  \ref{algorithm:otm_step}.

\begin{algorithm}
	\caption{Algorithmic implementation of a computation step in OTM}
	\begin{algorithmic} 
		\REQUIRE Initial nodal set and material point set\\
		1. Compute local mass matrix and local nodal force vector\\
		2. Update primary variables and nodal coordinates \\
		3. Update material point coordinates \\
		4. Constitutive updates at material point \\
		5. Search Algorithm to update  support domains \\
		6. Recompute shape functions\\
	\end{algorithmic}
	\label{algorithm:otm_step}
\end{algorithm}

\subsection{Update of Primary Variables}\label{subsec:primary_variable_update_otm_method}
As shown in \citep{2017_weissenfels_wriggers}, the OTM method can also be derived from the weak form and the formulation is made with respect to the current configuration, balancing the virtual work at the boundary with virtual work inside of the body and the inertia term.

\begin{equation} 
\int_{\Omega}\delta \mathbf{u}\cdot\rho \boldsymbol{\ddot{u}}dv\, +\, \int_{\Omega}\operatorname{grad}\delta\mathbf{u}:\boldsymbol{\sigma} dv\, =\, \int_{\Omega }\delta \mathbf{u}\rho\, \cdot \, \mathbf{\hat{b}}dv\, +\, \int_{\partial \Omega }\delta{\mathbf{u}}\, \cdot \, \mathbf{\hat{t}}da,
\label{eq:equation_of_motion}
\end{equation}
where the displacements $\mathbf{u}$ are the primary variables. The Cauchy stress tensor, specific body force and density correspond to $\boldsymbol{\sigma}$, $\mathbf{\hat{b}}$ and $\rho$ respectively. The surface traction $\mathbf{\hat{t}}$ is prescribed at the Neumann boundary.\medskip 

\begin{figure}[htb]
	\centering
	{\includegraphics[width=.50\textheight]{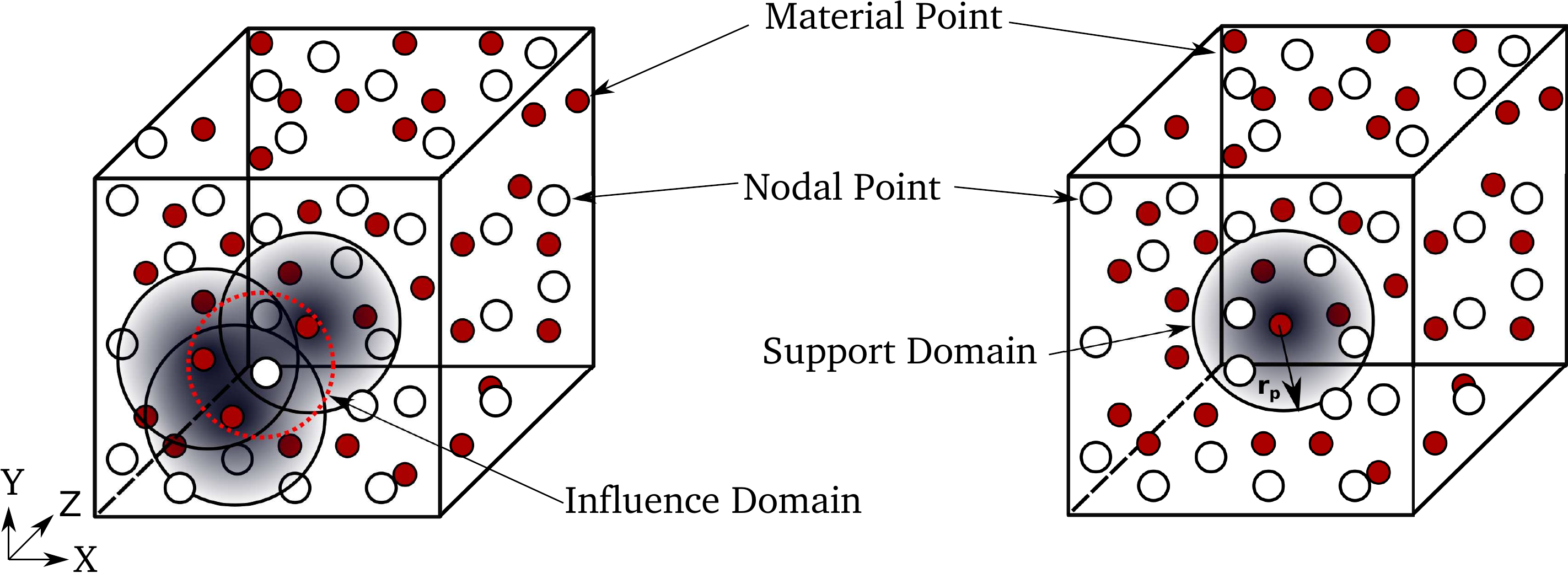}
	}\\
	\caption{Influence Domain of a node and Support Domain of a material point }
	\label{fig:support_and_influence}
\end{figure}

The support domain is defined as the domain around each material point, containing nearest nodes in its neighborhood (Fig \ref{fig:support_and_influence}). This domain is updated at every computation step by applying a suitable search algorithm. The shape functions $N_{I}\,(\mathbf{x}_{p\,n})$ are continuously updated as the OTM method has the usual structure of updated Lagrangian procedures. At each material point, the test function and the displacements are approximated through shape functions $N_{I}\,(\mathbf{x}_{p\,n})$ and nodal values within its support domain
\begin{equation}
\mathbf{u}_{p}(\mathbf{x}_{p\,n})\,= \, \sum_{I\,=\,1}^{n_{np}}N_{I}(\mathbf{x}_{p\,n})\mathbf{u}_{I},\; \delta \mathbf{u}_{p}\, =\, \sum_{I\,=\,1}^{n_{np}}N_{I}(\mathbf{x}_{p\,n})\delta \mathbf{u}_{I},\; \operatorname{grad}\delta \mathbf{u}_{p}\, =\, \sum_{I\,=\,1}^{n_{np}} \mathbf{B}_{I}(\mathbf{x}_{p\,n})\delta \mathbf{u}_{I},
\label{eq:material_point_displacements}
\end{equation}
where $n_{np}$ specifies the number of nodes in the support domain of each material point at current computation step. The matrix $\mathbf{B}_{I}(\mathbf{x}_{p\,n})$ contains the derivatives of shape functions at node $I$. In contrast to the FEM, overlapping of support domains is allowed in OTM method. In problems of large deformations, non-admissible nodal distributions can be eliminated by the update of support domains at every time or load step.

Using  \eqref{eq:material_point_displacements}, the approximation of  \eqref{eq:equation_of_motion} can be transformed into an algebraic equation using an assembly procedure 
\begin{equation} 
\left [ A_{p\,=\,1}^{n_{mp}}\sum_{I}^{n_{np}}\sum_{J}^{n_{np}}N_{I}(\mathbf{x}_{p})\mathbf{1}N_{J}(\mathbf{x}_{p})m_{p} \right ]\, \cdot \, \mathbf{\ddot{u}}\, =\,A_{p\,=\,1}^{n_{mp}}\sum_{I}^{n_{np}} \left [N_{I}(\mathbf{x}_{p})\mathbf{\hat{b}}_{p}m_{p}\, -\, \mathbf{B}_{I}(\mathbf{x}_{p})\mathbf{\sigma}_{p}v_{p}   \right	].
\label{eq:algebraic_form_eq_of_motion}
\end{equation}
where $\mathbf{\ddot{u}}$ is the global nodal acceleration vector, $n_{mp}$ is the total number of material points in the body, $m_{p}$ is the mass at the material point $p$ and $v_{p}$ is its volume in the current configuration.  In order to guarantee that the conservation of the mass during the computation, the mass of a material point is assumed to be constant.

Using the explicit central difference time integration scheme and the concept of lumped mass matrix, the equilibrium of the body is transformed into a set of independent nodal equilibrium equations. This step is equivalent to the Finite Element Method framework given in \citep{2006_bathe}, for instance
\begin{equation} 
m_{I\,n}\, \frac{\Delta \mathbf{u}_{I\,n+1}\, -\, \Delta \mathbf{u}_{I\,n}}{(\Delta t)^2}\, =\, \mathbf{p}_{I\,n} + \mathbf{r}_{I\,n}.
\label{eq:nodal_equilibrium_equation}
\end{equation}
The boundary forces, $\mathbf{p}_{I\,n}$, are prescribed only at the nodes of the Neumann boundary. The nodal residual vector $\mathbf{r}_{I\,n}$ and the nodal mass $m_{I\,n}$ are given as
\begin{equation}
m_{I\,n}\, =\, \sum_{p}^{n_{mp}^{I}}N_{I\,n}(\mathbf{x}_{p\,n})m_{p},\; \mathbf{r}_{I\, n}\, =\, \sum_{p}^{n_{mp}^{I}} \left [N_{I\, n}(\mathbf{x}_{p\,n})\mathbf{\hat{b}}_{p\,n}m_{p}\, -\, \mathbf{B}_{I}(\mathbf{x}_{p\,n})\mathbf{\sigma}_{p\,n}v_{p}   \right ]
\label{eq:nodal_mass_and_residue_formulation}
\end{equation}
where $n_{mp}^{I}$ is the number of material points in the influence domain of Node $I$. The corresponding material points within each influence domain can be determined from the support domain directly, without any need of additional search algorithm, see Fig \ref{fig:support_and_influence}.

By solving  \eqref{eq:nodal_equilibrium_equation}, the nodal position vector can be updated from the displacement increments of the next computation step
\begin{equation}
\mathbf{x}_{I\,n+1}\, =\, \mathbf{x}_{I\,n}\, +\, \Delta \mathbf{u}_{I\,n+1}
\label{eq:nodal_position_update}
\end{equation}

The integral is evaluated by a single point in each support domain. LME shape functions (see Section \ref{subsec:shape_otm_method}) are   rational exponential functions. Hence,                                                                                                                                                                                                                                                                                                                                                                                                                                                                                                   the total material points are not enough to accurately integrate the weak form in \eqref{eq:equation_of_motion}. A stabilization term is added to the nodal residual vector to penalize the inaccurate behavior due to underintegration within every support domain, see   \citep{2017_weissenfels_wriggers}: 
\begin{equation} 
r_{I\,n-stab}\, = \, r_{I\,n}\, -\, \varepsilon \sum_{p}^{n_{mp}^{I}}N_{I\,n}(\mathbf{x}_{p\,n})\mathbf{e}_{I,p\,n}
\label{eq:stabilization_term_addition_to_residue}
\end{equation}
where, $\varepsilon$ is the penalty parameter, $\mathbf{e}_{I,p\,n}$ is the error due to underintegration 
%
\begin{equation}
\mathbf{e}_{I,p\,n}\, =\, \frac{\mathbf{x}_{I\,n}\, -\,\mathbf{x}_{p\,n}\, -\, (\mathbf{\tilde{x}}_{I\,n}\, -\mathbf{\tilde{x}}_{p\,n}) }{\left \|\mathbf{x}_{I\,n- 1}\, -\, \mathbf{x}_{p\,n - 1}  \right \|},\qquad \mathbf{\tilde{x}}_{I\,n}\, -\, \mathbf{\tilde{x}}_{p\,n}\, =\, \Delta \mathbf{F}_{p\, n}\left [\mathbf{x}_{I\,n-1}\, -\, \mathbf{x}_{p\,n-1}  \right ]
\label{eq:underintegration_error_and_dist_vector}
\end{equation}
and $\Delta \mathbf{F}_{p\, n}$ is the increment of the deformation gradient, as computed in  \eqref{deformation_grad_update}.


\subsection{Update of Kinematic Quantities}\label{subsec:mp_updates_otm_method}
The position vector of a material point at the next computation step is updated by multiplying the shape functions at the current time step with the nodal coordinates of the next time step
\begin{equation}
\mathbf{x}_{p\,n + 1}\, =\, \sum_{I}^{n_{np}^{n}}N_{I\, n}(\mathbf{x}_{p\, n})\mathbf{x}_{I\, n+1}
\label{mat_pt_coordniate_equation}
\end{equation}

The deformation gradient  at the next computation step, $n+1$, 

\begin{equation} 
\mathbf{F}_{p\, n+1}\, =\, \Delta \mathbf{F}_{p\, n+1}\mathbf{F}_{p\, n}
\label{deformation_grad_update}
\end{equation}
is updated in terms of the current value of the deformation gradient, $\mathbf{F}_{p\,n}$, and the increment of the deformation gradient is 

\begin{equation}
  \Delta \mathbf{F}_{p\, n+1}\, =\, \mathbf{1}\, +\, \sum_{I}^{n_{np}^{n}}\frac{\partial N_{I\, n}(\mathbf{x}_{p\, n})}{\partial \mathbf{x}}\Delta \mathbf{u}_{I\, n+1}\, ,
 \label{deformation_grad_increment}
\end{equation}

Accordingly, the volume and density at each material point is also updated by $\Delta \mathbf{F}_{p\, n+1}$

\begin{eqnarray}
v_{p\, n+1}\, &=& \, \textup{det}(\Delta \mathbf{F}_{p\, n+1})v_{p\, n} \\
\rho^{n+1}\, & = & \, \frac{m_{p}}{v_{p\, n+1}}
\label{volume_and_density_mat_pt}
\end{eqnarray}

\subsection{Local Max-Ent shape functions}\label{subsec:shape_otm_method}

In meshfree methods, the polynomial basis functions, which are normally used within the finite element framework, are not appropriate. In OTM method, local maximum entropy (LME)  \citep{2006_arroyo_ortiz} approximation function is used which has to be determined for an arbitrary number of nodes within the support domain. The LME shape functions possess weak Kronecker-$\delta$ property at the boundary and it is fulfilled only on convex boundaries, see   \citep{2010_li_habbal_ortiz}. Also, the LME shape functions does not fulfill either the first order completeness or the partition of unity condition. In order to achieve convergence to the correct solution of the equation of motion, computational algorithms should fulfill these basic conditions, see   \citep{hughes_fem} and   \citep{1998_Belytschko_Krongauz_Dolbow_Gerlach}. \medskip

The LME shape functions has an exponential ansatz and it belongs to the class of radial basis functions. First order completeness condition is enforced using Lagrangian multiplier method and the partition of unity condition is enforced through normalization 
\begin{equation}
N_{I}(\mathbf{x}_{p})\, =\, \frac{Z_{I}(\mathbf{x}_{p})}{Z},\qquad Z_{I}\, =\, \exp{-\beta \left | \mathbf{x}_{p}\, -\,{} \mathbf{x}_{I}  \right |^{2}\, +\, \lambda (\mathbf{x}_{p}\, -\, \mathbf{x}_{I})},\qquad Z\, =\, \sum_{I}^{n_{np}}Z_{I}
\label{lme_shape_function}
\end{equation}
where $\lambda$ is a Lagrangian multiplier, which is determined by solving $\sum N_I(x_p)(x_p-x_I)=0$  using Newton- Raphson algorithm. The parameter $\beta$ is calculated as $\beta \, =\, \frac{\gamma }{h^{2}}$, where $\gamma$ controls the degree of locality of LME shape functions and it should be in the range of 0.8 to 4, and $h$ is the characteristic nodal spacing. 


\section{Software Design}\label{sec:MPI_intro}

The \PG{method} is written for use on multi-CPU architectures. The parallel codes are written in C++ (would also be possible with Fortran) and make use of its object-oriented features. The \PG{code} utilizes the Message Passing Interface (MPI) for communication and synchronization between processes. MPI is a standard paradigm for implementing parallel  \PG{software} in distributed memory platforms, \citep{balaji_buntinas_goodell_et_al_2010,plimpton_devine_2011}. In order to exchange message and manage processes, MPI provides  a collective set of library routines. It is generally used in high end computing applications involving intensive calculations \citep{notay_napov_2015}. \medskip

The  approach to parallelize the OTM method with MPI is to separate the spatial domain into distinct subdomains and allocate  nodes and material points to each MPI process, such that each process treats its own subdomain independently. One advantage of this approach is the minimum impact on the contents and structure of a serial code. Halo regions  of nodes and material points are then distributed between the subdomains at every computation step such that the primary nodal variables and constitutive updates at the material points  can be computed in parallel.


\subsection{Domain Decomposition}\label{subsec:domain_decomposition}

\begin{figure}[htb]
	\centering
	{\includegraphics[width=.30\textheight]{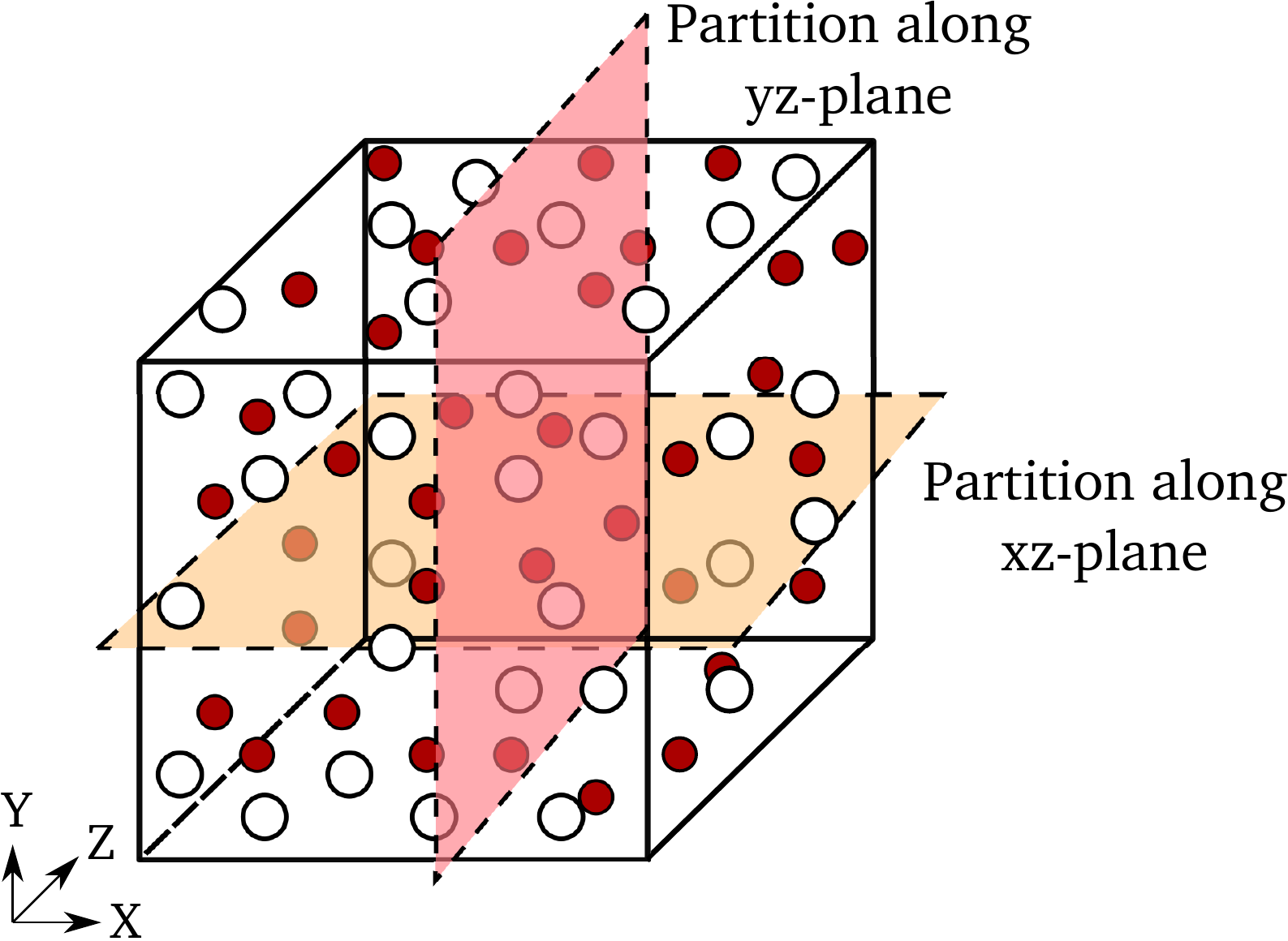}
	}\\
	\caption{Domain decomposition into four processes using RCB. }
	\label{fig:domain_decomposition}
\end{figure}

To decompose the domain, the Recursive Coordinate Bisection (RCB) algorithm is used \PG{from Zoltan library \citep{ZoltanDevelopersGuideV3}.} 
The objective of the partitioning library is to provide a  initial computational workload which  is uniformly distributed. This is accomplished by a distribution of almost equal number of particles (nodes and material points) in each process. Domain decomposition is conducted by cutting along the partition planes in the spatial domain recursively (Fig \ref{fig:domain_decomposition}). Each sub-domain is assigned to one process. Hence, the decomposition  depends on the number of processes and the domain size \citep{selvam_hoffmann_2015}. The goal of using this domain decomposition algorithm is to ensures geometrical locality of the particles and to simplify the creation of halo regions. Both nodes and material points carry their influence and support domain information respectively during the distribution process.  



Movement of particles and subsequent adjustments in the subdomain will cause load-imbalance among the processes. The computational load at a given time interval is monitored to re-assign the workload evenly among the processes and to minimize the communication. 
Major constraints are minimizing the computational efforts to compute the new division of the domain and minimizing the number of particles that need to be migrated among the processes.   For the purpose of dynamic load balancing, the Recursive Coordinate Bisection algorithm is called. At optimized time intervals, within the mid-increment of the time step, to update the subdomain boundaries if required.

\subsection{Dynamic Halo Regions}\label{subsec:communication}

\begin{figure}[htb]
	\centering
	{\includegraphics[width=.40\textwidth]{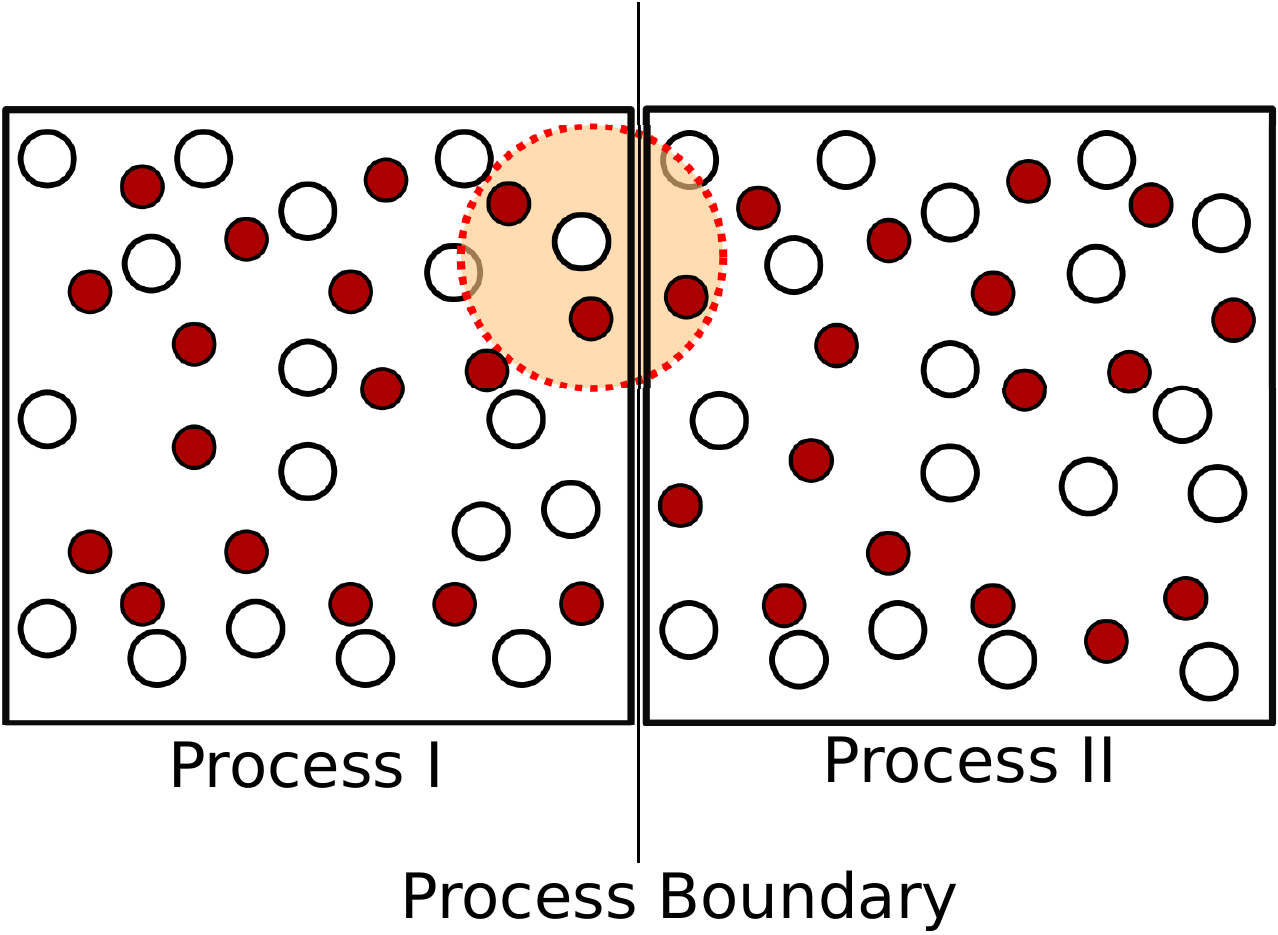}
	}\\
	\caption{Influence domain of a node spread across multiple processes. }
	\label{fig:influence_domain_across_processes}
\end{figure}

The nodal and material point updates are performed in each subdomain in parallel. For a node and material point, its influence and support domain could be spread across multiple processes (Fig  \ref{fig:influence_domain_across_processes}). Nodes and material points, which are close to the division boundaries of subdomains need to share information. For this, the halo regions are necessary. These halo regions are copies of nodal and material point data that are sent to neighbor processes via two communication steps. \medskip   

\begin{figure}[htb]
	\centering
	{\includegraphics[width=.40\textwidth]{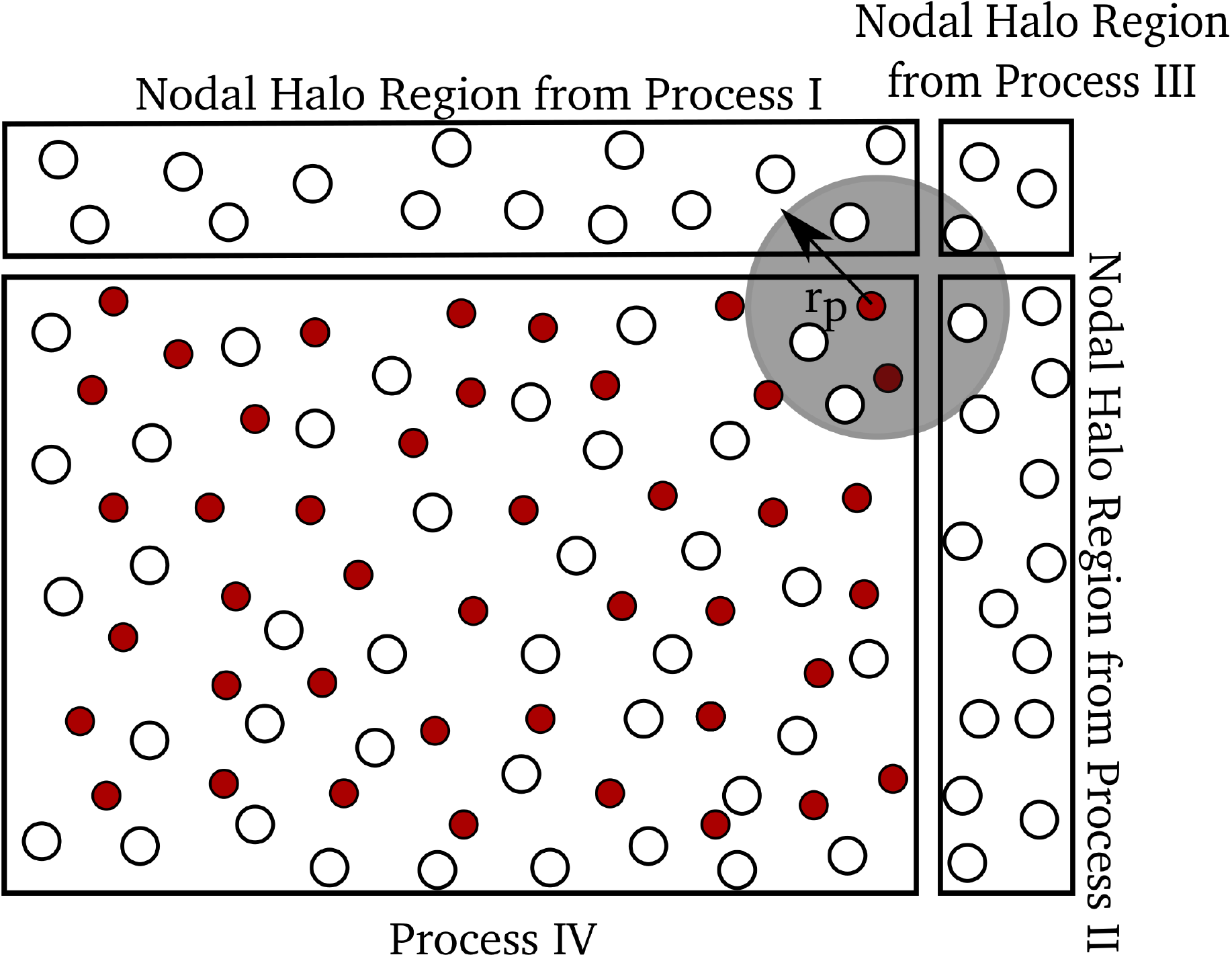}
	}\\
	\caption{ Nodal halo region and support domain with support nodes from halo region. }
	\label{fig:nodal_halo_with_support_domain}
\end{figure}

\begin{figure}[htb!]
	\centering
	{\includegraphics[width=.40\textwidth]{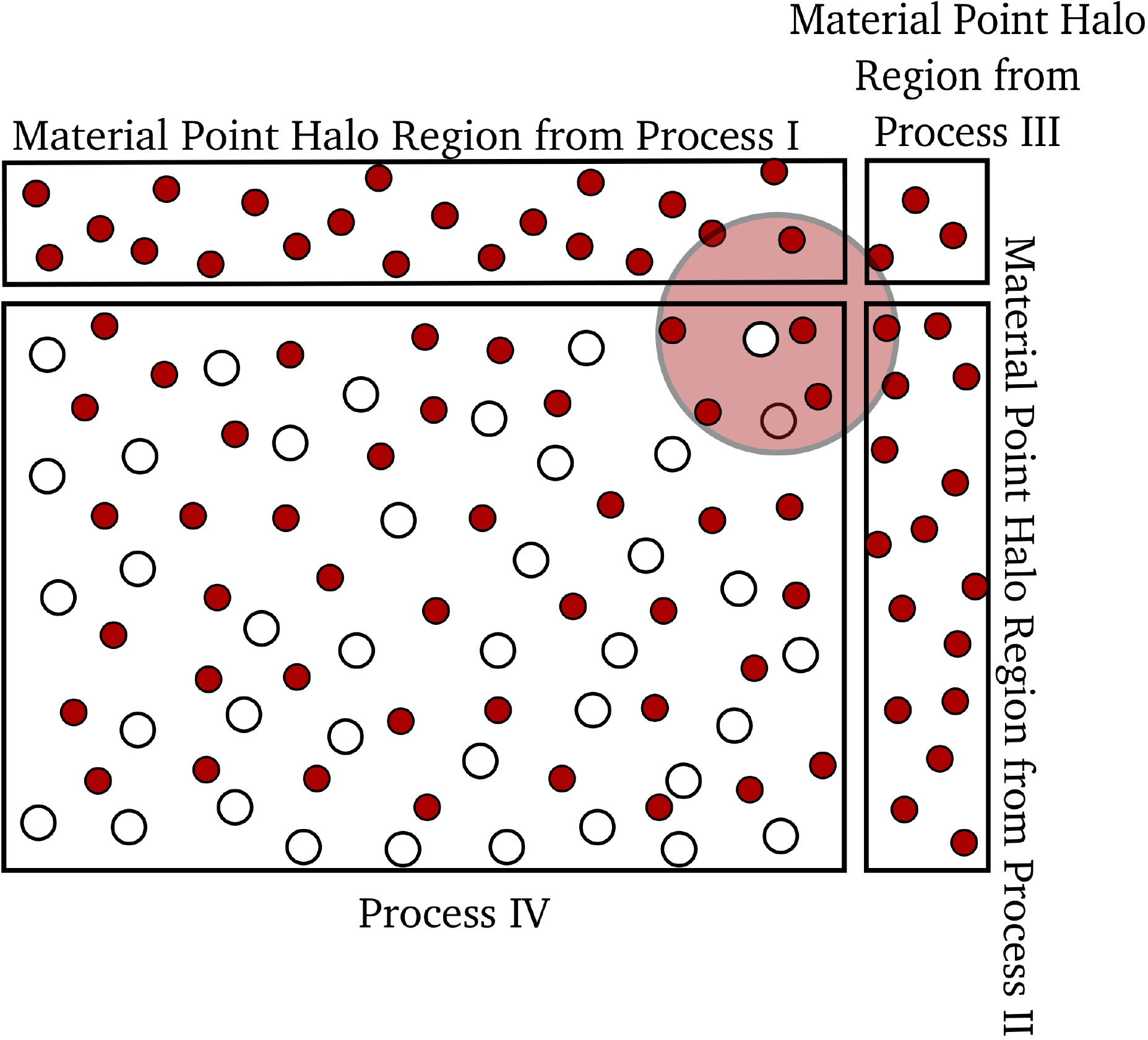}
	}\\
	\caption{Material point halo region and influence domain using material points from halo region. }
	\label{fig:material_point_halo_with_influence_domain}
\end{figure}

 After nodal and material point updates, the halo regions are constructed dynamically during the computation, depending on the amount of communication. The first round  involves nodal halo communication for the material point updates, where position, velocity, influence domain and other nodal data are communicated.  Afterward, all data  at material points can be computed within each subdomain. For material points whose support nodes are located in neighboring subdomains, its support domain is reconstructed through halo nodes (Fig \ref{fig:nodal_halo_with_support_domain}). Hence, support domains are formed using nodes in its own subdomain and nodal halo region. After the update of material point information, the second round of communication involves material point halo communication for the nodal updates. Similarly,  influence domain of nodes at the boundary of subdomains are  reconstructed through material point halo regions (Fig \ref{fig:material_point_halo_with_influence_domain}). Nodal updates take place locally at each subdomain using the information from its own subdomain and from material point halo region. \medskip

\begin{figure}[htb]
	\centering
	{\includegraphics[width=.40\textwidth]{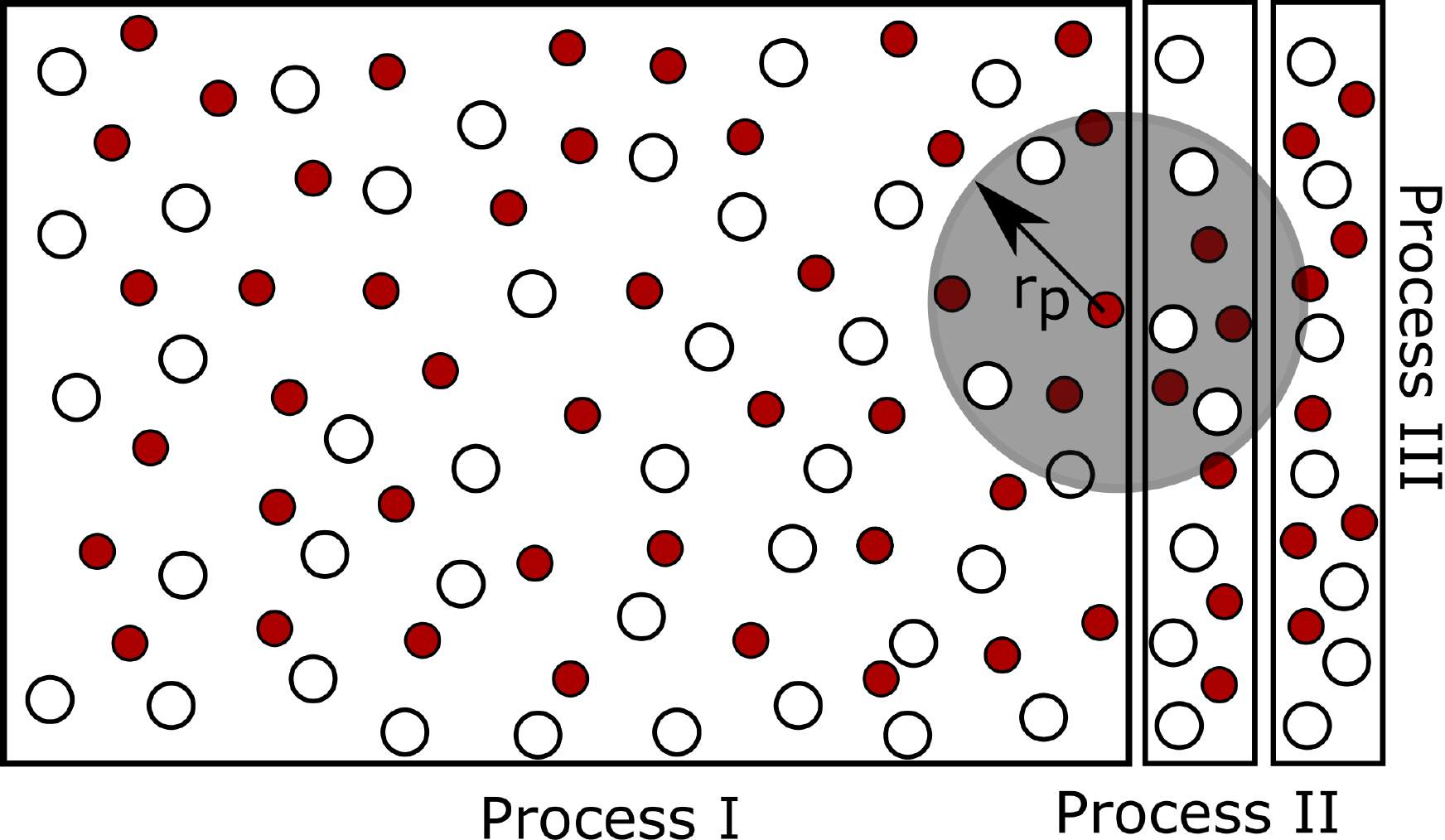}
	}\\
	\caption{Support domain across multiple processes. }
	\label{fig:support_across_multiple_processes}
\end{figure}

\begin{figure}[htb]
 	\centering
 	{\includegraphics[width=.25\textwidth]{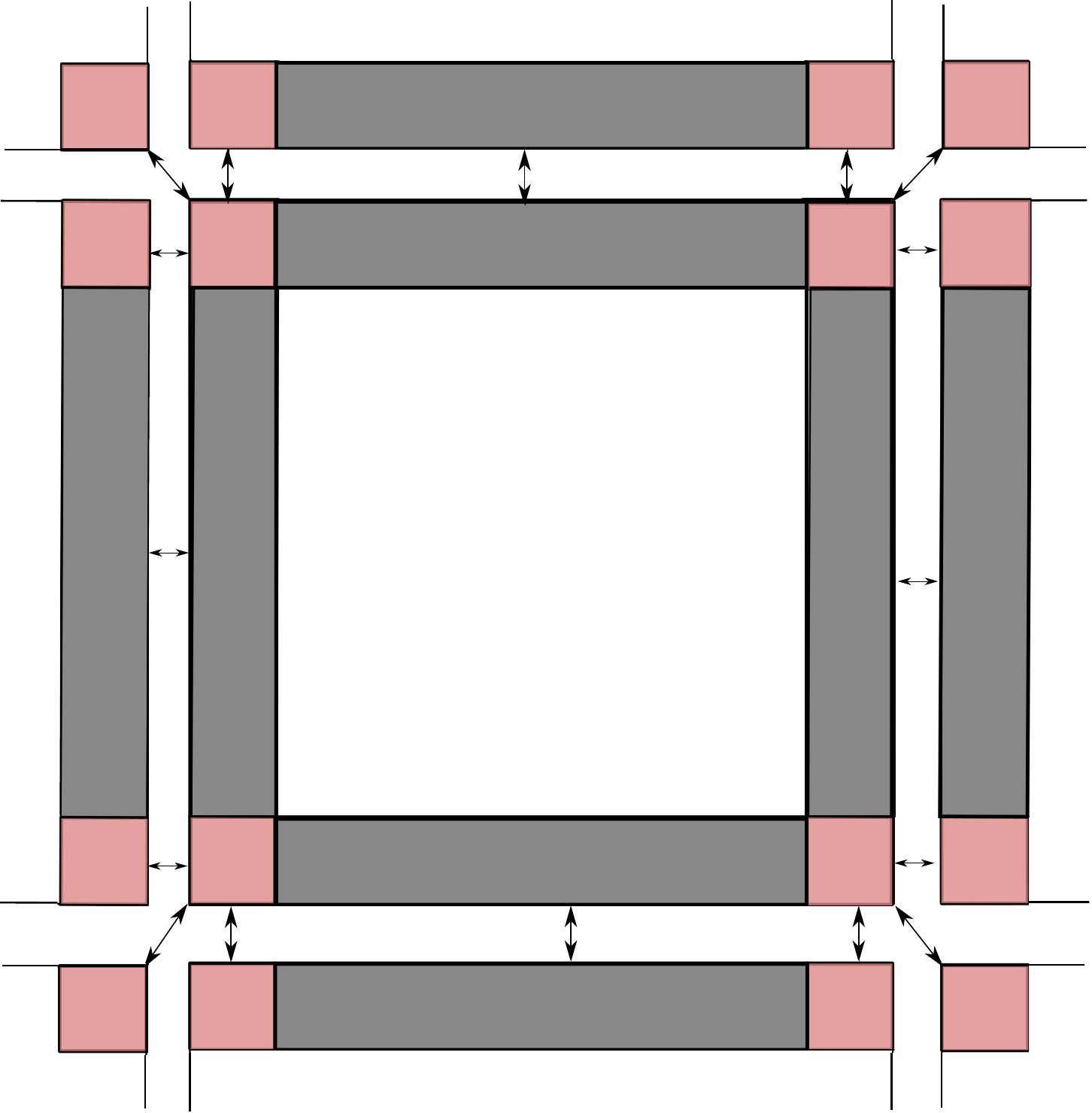}
 	}\\
 	\caption{Schematic Representation of Nearest Neighborhood Communication for a typical halo region update operation, showing the halo exchanges for the faces and corners of sub-domains: Grey regions represent the halo communication among the faces of the sub-domains (each with one neighbor) and red regions represent the halo exchange among the corner parts of the sub-domains (each with three neighbors).}
 	\label{fig:nearest_neighborhood_comm}
 \end{figure}

\begin{figure}[htb]
 	\centering
 	{\includegraphics[width=.40\textwidth]{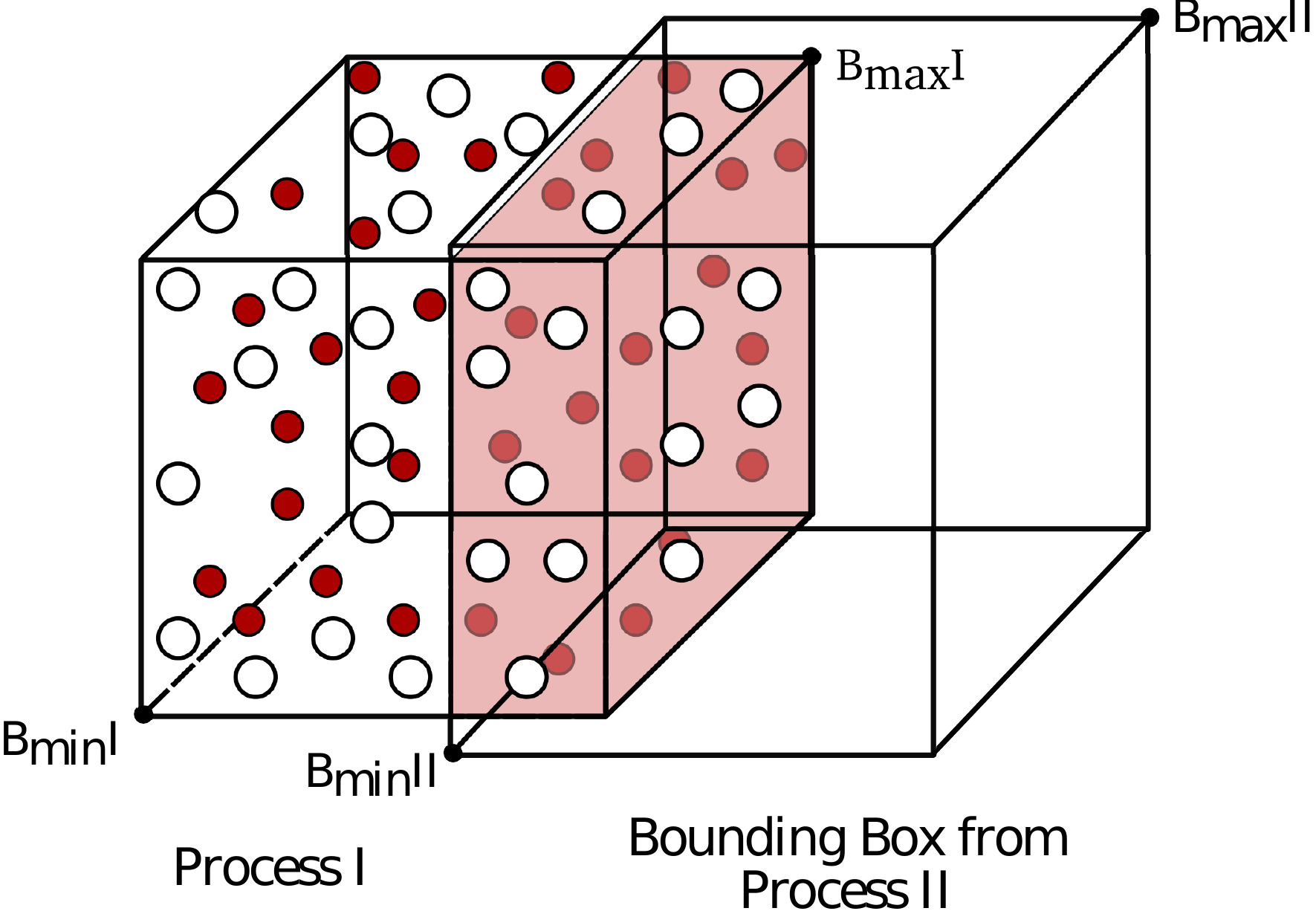}
 	}\\
 	\caption{Bounding Box intersection between Process I and Process II and Identification of halo nodes in Process I in the intersection region is depicted in the Figure. Nodes in the intersection region will be sent as halo nodes to Process II.}
 	\label{fig:bounding_box_intersection}
 \end{figure}

At the beginning of the halo communication process, the neighbors of each subdomain need to be detected. This will enable every subdomain to initiate communication locally only with its neighbors and to implement the nearest neighbor communication patterns (\textit{sparse collective operations}). In general, for large scale applications, efficient implementation of sparse collective communication operations is most important  \citep{hoefler_traff_2009}. At every time step, identification of nearest neighbors of a subdomain is facilitated through the  process of bounding box intersection with its neighbor subdomains. A set of local neighborhoods (\textit{process neighborhood}) is defined for every subdomain (Fig \ref{fig:nearest_neighborhood_comm}). Each process neighborhood consists a list of \textit{k} target processes and a list of \textit{k} source processes. For each subdomain, halo regions will be sent to target processes and simultaneously, it will receive halo regions from the same target processes. So, the source and target processes are same for each process but the amount of information to be received and sent may differ. Bounding box consists of coordinate information of both nodes and material points located at the lower and upper bounds of each subdomain and it is recomputed at every computation step. \PG{Even though the problem never occurred in our simulations, it is always checked that the minimal dimension of the bounding box is larger than the radius of the support domains so that only the nearest neighbors need to be detected, see Figure~\ref{fig:support_across_multiple_processes} for an illustration of the situation to be avoided. If such situation occurred, load balancing should be  realized by calling Zoltan library in order to adapt the domain decomposition and transfer particles between subdomains}.  In Fig \ref{fig:bounding_box_intersection}, $(B_{max}^{II}, B_{min}^{II})$ represents the bounding box of Process $II$ and $(B_{max}^I, B_{min}^I)$ represents the bounding box of Process $I$. Before performing intersection, bounding boxes from neighboring processes are extended by a width equivalent to maximum support radius of the sub-domain. \PG{Also, it ensures that there are no missing neighbor detection.} This is performed at \PG{regular intervals} to determine the extent of overlap of bounding boxes, i.e. halo regions. The maximum support radius at each sub-domain is used to extend the bounding boxes gathered from neighboring processes. Width of halo region for each sub-domain is identified as the overlap region between the bounding boxes of each sub-domain. \medskip

\begin{figure}[htb]
 	\centering
 	{\includegraphics[width=.95\textwidth]{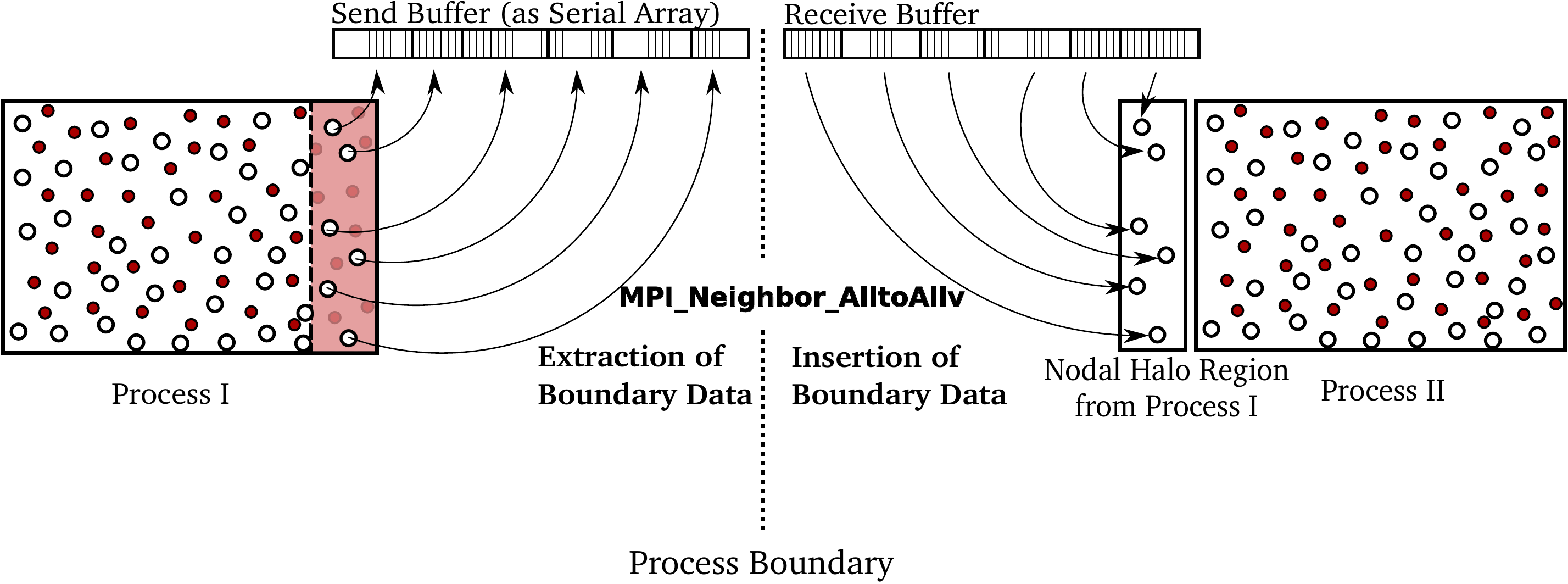}
 	}\\
 	\caption{MPI Communication from Process I to Process II }
 	\label{fig:MPI_comm}
 \end{figure}

 Nodes within every intersecting bounding boxes are identified as halo nodes, see Fig \ref{fig:bounding_box_intersection}. Halo nodes, which are to be sent, are serialized,i.e., the nodal data is represented as a large array of chars and stored in a buffer. Halo nodes sharing boundary with each neighbor is sent to the specific neighboring sub-domain (Fig \ref{fig:MPI_comm}). Here, MPI virtual topology functionality is used for sparse collective operations, it uses the set of local neighborhoods, i.e., source and target lists. Graph topology interface is used as it provides full flexibility in describing neighborhoods and the communication graphs are not limited to symmetric exchange patterns, which is in contrast to the Cartesian topology mechanism \citep{mpi_standard_2012}. Before sending the halo nodes, the processes, at first, communicate how many nodes are to be exchanged along with  total size of nodal data. After determining the total size of each buffer, memory allocation is made in the target process where the halo region is to be received (receive buffer) from its neighbor ([Fig \ref{fig:MPI_comm}). The overall nodal halo communication step is sketched in Algorithm  \ref{algorithm:nodal_halo_step}.

 \begin{algorithm}
 	\caption{Nodal Halo Communication Step}
 	\begin{algorithmic} 
 		\REQUIRE Bounding box computation at every process \\
 		\REQUIRE Detection of neighbor processes \\
 		1. Exchange bounding box with all neighbor processes using MPI\_Allgather\\
 		2. Identify intersecting bounding boxes \\
 		3. Identify nodes at the intersection (or Halo nodes) to be sent \\
    	\REQUIRE Create local process neighborhood using MPI\_Dist\_graph\_create\_adjacent \\
    	1. Determine the nodal size for halo. \\
    	2. Exchange the nodal sizes with nearest neighbors using MPI\_Neighbor\_alltoallv \\
    	3. After receiving the nodal sizes at destination process, allocate memory for receive buffer \\
    	4. Pack the halo nodes to be sent in a send buffer. \\
    	5. Exchange the halo node information using  MPI\_Neighbor\_alltoallv.
   	\end{algorithmic}
   \label{algorithm:nodal_halo_step}
 \end{algorithm}

\begin{figure}[htb]
	\centering
	{\includegraphics[width=.40\textwidth]{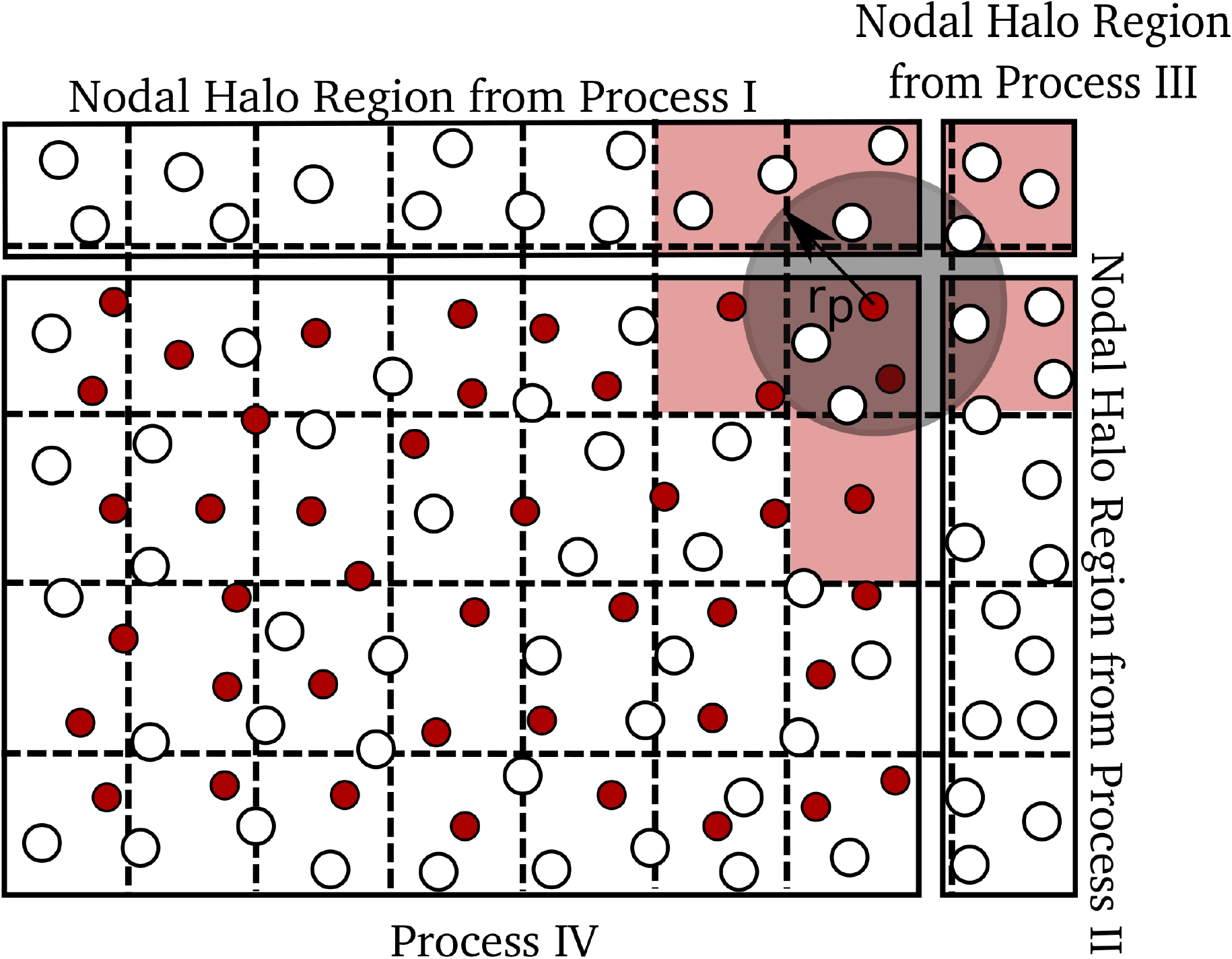}
	}\\
	\caption{Linked Cell Method }
	
	\label{fig:nodal_halo_with_linked_cell}
\end{figure}

 After the nodal halo communication, material points perform the search process  using nodal information from its own subdomain and from nodal halo region. In order to improve the computational efficiency of search algorithms, the linked cell method  \citep{griebel_knapek_zumbusch_2007} has been implemented. Linked cell method within OTM method is a new feature in this work. In solids undergoing large deformations or in fluid simulations, both the nodes and material points may change its position with time. An efficient search algorithm is needed to dynamically update the support domain while solving the equations without incurring excessive computational costs. Linked cell method significantly reduces the computational efforts, when the number of particles is large. The main idea of the linked list is to map the nodal positions on a grid. All the nodes have a unique particle ID and a data structure stores all the information of each grid. For each cell, a list of  nodal IDs and pointer to those nodes are stored. After the nodal updates and subsequent formation of nodal halo regions, both the subdomain and nodal halo region is subdivided into static cells, (Fig \ref{fig:nodal_halo_with_linked_cell}). Only nodes in the vicinity of a material point are checked during the search process. This is done by identifying the cells which intersect with the search radius and nodes in those intersecting cells are considered within the search process. A brief sketch of the steps involved is shown in Algorithm \ref{algorithm:linked_cell}. \medskip
 

 \begin{algorithm}[htb]
 	\caption{Algorithmic scheme for Linked Cell Method}
 	\begin{algorithmic} 
	
 		\REQUIRE Nodal Halo Region \\
 		\REQUIRE New support radius  \\
 		1. Divide the subdomain and nodal halo region into cells, see Fig  \ref{fig:nodal_halo_with_linked_cell}\\
 		2. Identify intersection of cells with the support domain \\
 		3. Find the nodes which belong to the material point \\
 	\end{algorithmic}
 	\label{algorithm:linked_cell}
 \end{algorithm}


 After material point updates, communication of material point halo region follow in the same line as of nodal halo communication, starting with the bounding box intersection, communicating total number of material points to be exchanged, and communication of material point information. \medskip

\begin{algorithm}[htb]
	\caption{Parallel OTM Time Step}
	\begin{algorithmic} 
		\STATE For Process $\mathbb{P}^{I}\,,I = 1,....,P:$ \\
		\REQUIRE Reading of Input information and Process $\mathbb{P}^{I}$ storing its own set of nodes and material points.
		\begin{itemize}
			\item Initial nodal set and material point set
			\item Initial support domain of material points
		\end{itemize}
	    \REQUIRE Domain Decomposition by Zoltan,  see Fig  \ref{fig:domain_decomposition}.
	    \REQUIRE Initial material point halo regions(steps are similar to Algorithm  \ref{algorithm:nodal_halo_step})
	    \STATE For computation step $t_{k}\,\rightarrow\,t_{k\,+\,1}$\\
	    1. Complete the influence domain with halo material points. \\
		2. Compute the local mass matrix and local nodal force vector.\\
		3. Update primary variables and nodal coordinates  \\
	  \REQUIRE Nodal halo regions (for details, see Algorithm  \ref{algorithm:nodal_halo_step}) \\
	  \REQUIRE Load balancing at optimized intervals, let's say at every time increments of $t_{k\,+\,500}$
	  \begin{itemize}
	  	\item Clear both nodal and material point halo regions
	  	\item Call Zoltan functions for load balancing (steps are similar to domain decomposition as in Section  \ref{subsec:domain_decomposition})
	  \end{itemize}
		4. Complete the support domain with halo nodes, see Fig  \ref{fig:nodal_halo_with_support_domain}.  \\
		5. Update material point coordinates.\\
		6. Constitutive updates at material point. \\
		7. Division of subdomain and nodal halo region into cells (Linked Cell Method, see Fig  \ref{fig:nodal_halo_with_linked_cell} )\\
		8. Search algorithm to update the support domains\\
		9. Recompute shape functions\\
	   \REQUIRE Material Point halo regions (steps are similar to Algorithm  \ref{algorithm:nodal_halo_step})	
	\end{algorithmic}
	\label{algorithm:parallel_otm_step}
\end{algorithm}
 
 \subsection{Data Management Strategies}
 \label{subsec:data_management_strategies}
 The core of parallel OTM method  is formed by data structures and algorithms implemented as C++ templates (Not limited to C++ and could be also implemented in Fortran). To store all the data of the nodes and material points, C++ classes have been defined. For the management of nodal and material point data (removal or addition),  STL maps to store pointers to objects of nodal and material point data are preferred. This gives us flexibility for quicker removal and addition of nodal and material point data during load-balancing and for the formation of support and influence domains. \medskip
 
 Information that is contained in a particle (node or material point) are its identifier (Global particle ID), coordinates, flags (indicators, such as,  the particle is a node or material point and if the node is on the physical boundary of the problem domain) and its affiliate (rank of the process that the node or material point belongs to). Additional information that is contained in a node and material point is its pointer-based influence and support domain information respectively. 
 With the help of this data structure, every subdomain handles pointers to objects of nodes and material points, bounding box information (maximum and minimum coordinates), and neighbor information (halo regions for nodes and material points). Choosing a proper way to handle this STL container depends on the problem itself. For instance, there is continuous update of support and influence domain in the OTM method, whose sizes can vary dynamically at every time step. After optimized intervals of dynamic load-balancing, the pointers to  new particles (nodes and material points) are handled effectively by this container.\medskip
 
 For nodal and material point halo communication, the data is packed into a serial array, whose size is varying. So, flexible data structures have been designed to pack all the information in the  buffer. The message size for each node or material point is maintained as number of nodal or material point variables multiplied with the (size of double precision floating number), in order to prevent any kind of memory misalignment issues while packing information of mixed data types in a serial array. \medskip
 
 The size of each nodal information is of arbitrary number of bytes due to the varying size of its influence domain. 
  In \citep{2014_li_stalzer_ortiz},  MPI data structure  was used to pack the nodal information. This restricts the information to be packed since only fixed-size information could be used to communicate. Here,  the  influence domain information of a node is packed more efficiently using a flexible size for every node. Similarly, for every halo material point, its support domain information is also included in the halo region. Packing support and influence information in halo region assists in localised updates within a subdomain.  For instance, whenever the support domains of boundary material points are updated (Step 8 of Algorithm  \ref{algorithm:parallel_otm_step}) and those material points are exchanged through halo communication, the updated support domain information of halo material points will assist in updating the influence domain of the nodes locally at each sub-domain. This flexibility feature for packing any amount of information for halo communication is necessary for localized updates of nodes and material points. \medskip


 Another advantage of using STL map for support and influence domain is that pointers to the support nodes or influence material points can be released while preserving their IDs. This proved to be helpful in situations where the support domains need to be constructed again using halo nodes after nodal updates. For instance, at time step $t_{k}$, support domains are updated (Step~8 of Algorithm~\ref{algorithm:parallel_otm_step}). Subsequently, for the material point updates (Steps 4-7 of Algorithm  \ref{algorithm:parallel_otm_step}) at time step $t_{k\,+\,1}$, the support domain computed at previous time step $t_{k}$ will be used. \medskip
 
 
 
  
  Object-oriented implementation, robustness and flexibility of the parallel \PG{method} to include additional physical phenomena are taken into consideration. With the use of \textit{Eigen} templated library, all the vector and matrix information are stored in contiguous memory locations and matrix operations are optimized. 
 

 \section{Parallel Performance} \label{sec:parallel_performance}
 
 The objective of parallelism is to perform simulation of larger and complex problems. To evaluate the ability of the parallel \PG{implementation}, strong scaling tests are conducted which measure the  performance with increasing number of processes, keeping the problem size constant.  \medskip
 
 \PG{The computation being explicit in time with lumped mass, we expect the nodal and material point updates to scale perfectly with the number of subdomains. The use of linked cells method makes the search for support domains independent on the number of processes. The number of neighbors is at most 8 (4 edges, 4 vertices) in 2D and 26 in 3D (6 faces, 12 edges, 8 vertices), so the neighbor-communications do not depend on the total number of processes. The number of particles in the halos decreases sublinearly with the number of processors, and the ratio between inner and halo particles strongly depends on the dimensionality of the physical space. So that at some point, the cost of forming the halo should increase in comparison with the cost of the update of inner particles, for increasing number of subdomains. To sum up, we can expect our implementation to scale well if we take a low-rank parallel reference, up to a sufficiently large amount of subdomains when the formation of halo regions dominates the computational cost.}
 \medskip
 
Every simulation is run for 2000 time steps. Variation of the computational efforts could occur between simulations due to fluctuations in cluster load and differences in configurations of the cluster nodes. Hence, each simulation is run for 3 times and the average CPU time is used in the studies. Output files are written in binary format of \textit{vtk} for every process. Time taken to write the data files is also taken into consideration. The computational time is the maximum wallclock time for a single time step in Algorithm  \ref{algorithm:parallel_otm_step}. Speedup is measured as 
\begin{equation}
\text{Speedup} \,=\, \frac{t_{n}}{t_{p}}
\label{eq:speedup_formula}
\end{equation}
For the baseline calculation, the sequential time \PG{$n = 1$} is used and  $t_{p}$ is the maximum wallclock time for a single time step with $p \geq n$. 
Efficiency is measured as

\begin{equation}
\text{Efficiency} \,=\, \frac{n\, \times \,t_{n}}{p\, \times \,t_{p}} \,=\,\text{Speedup}\, \times \, \frac{n}{p}
\label{eq:efficiency_formula}
\end{equation}


In this section, we will assess the strong scalability characteristics of our parallel approach. The  studies are performed on the LUIS Cluster of Leibniz Universität Hannover using only Haswell-based nodes. Each Haswell-based node consists of two 8-core Intel Xeon E5-2630 processors. All nodes are interconnected with the Infiniband technology. \PG{The cluster nodes are based on Torque/ Maui (qsub, qstat) workload manager.} Each sub-domain is assigned to one process (core).


 \subsection{Application to Taylor rod impact}\label{subsec:taylor_rod}
 The Taylor rod impact test is a widely accepted benchmark where a copper rod hits a rigid frictionless wall. The three-dimensional bar has a length of $L\,=\,32.4$~mm and a circular cross-section with radius $R_{0}\,=\,3.2$~mm. The initial velocity is 227~m/s.  \medskip

 \begin{figure}[htb]
 	\centering
 	{\includegraphics[width=.30\textheight]{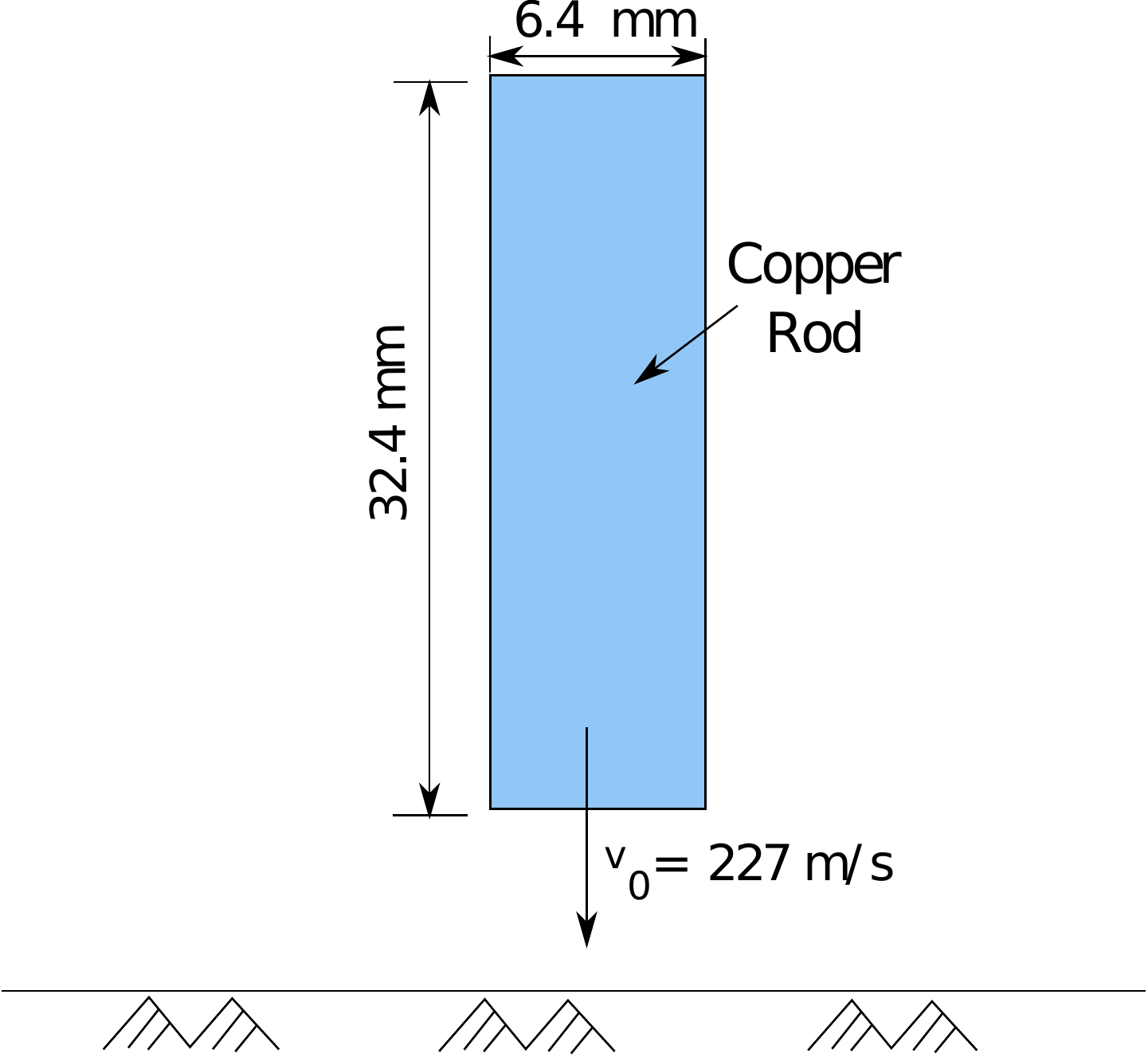}
 	}\\
 	\caption{Geometrical setup of the Taylor rod test. }
 	\label{fig:taylor_rod}
 \end{figure}
 
 \subsubsection{Material Model}
 In this benchmark problem, a finite plasticity material model with linear isotropic hardening is  used to  model the  behavior of the rod. The formulation is based on the multiplicative split of the deformation gradient into an elastic and plastic part
 \begin{equation}
 \textbf{F}_{p\,n}\, =\, \textbf{F}^{e}_{p\,n}\textbf{F}^{p}_{p\,n}
 \end{equation}
 
 Assuming the strain behavior as Hencky strain measure, it can be expressed in terms of the left elastic Cauchy-Green strain tensor $\textbf{b}^{e}_{pn}$ as
 \begin{equation}
 \boldsymbol{\varepsilon}^{e}_{p\,n}\, =\, \textup{ln} \textbf{V}^{e}_{p\,n}\, =\, \textup{ln}\,\left ( \sqrt{\textbf{b}^{e}_{p\,n}}\right )
 \label{eq:hencky_strain}
 \end{equation}
 where $\textbf{V}^{e}_{p\,n}$ is the elastic left stretch tensor.\medskip
 
 Using the exponential map integrator, the Equation  \ref{eq:hencky_strain} can be expressed as
 \begin{equation}
 \boldsymbol{\varepsilon}^{e}_{p\,n}\, =\, \textup{ln} \left ( \sqrt{\bar{\textbf{b}}^{e\, tr}_{p\,n}} \right )\, -\, \frac{\gamma_{p\,n}\, -\,\gamma_{p\,n-1} }{\Delta t}\frac{\partial f_{p\,n}}{\partial \boldsymbol{\tau}_{p\,n}}, \: \bar{\textbf{b}}^{e\, tr}_{p\,n} \, =\, \textbf{Q}_{p\,n}\textbf{b}^{e\, tr}_{p\,n}\textbf{Q}_{p\,n}^{T}
 \label{eq:hencky_strain_and_rotation_tensors}
 \end{equation}
 The elastic left Cauchy Green tensor is transformed in the principal stress space using the rotation tensor~$\textbf{Q}_{pn}$.\medskip
  
 The onset of plastic yielding is defined by the yield function $f$. The yield surface divides the elastic domain from the plastic domain and the Kirchoff stresses $\boldsymbol{\tau}_{pn}$ must lie within the elastic domain or on the yield surface. von Mises plasticity model is used and its deviatoric part leads to plastic deformations
 \begin{equation}
 \boldsymbol{\tau}_{p\,n}\, =\, p\mathbf{I}\, +\, \mathbf{s}_{p\,n}\, =\, K \operatorname{tr}( \boldsymbol{\varepsilon}^{e}_{p\,n})\mathbf{1}\, +\, 2\mu \left (\boldsymbol{\varepsilon}^{e}_{p\,n}\, -\, \frac{1}{3}\boldsymbol{\varepsilon}^{e}_{p\,n}\cdot \mathbf{1} \otimes  \mathbf{1}  \right )
 \end{equation}
 where, the constants $K$ and $\mu$ are the compression modulus and the second Lam\'{e} constant.

 
 The plastic flow (or evolution of the plastic deformation gradient) can be defined in terms of the plastic strain as
 \begin{equation}
\textbf{d}^{p}_{p\,n}\, =\, \dot{\gamma}_{p\,n} \frac{\partial f_{p\,n}}{\partial \boldsymbol{\tau}_{p\,n}}
  \end{equation}
where $f_{p\,n}$ is the yield function and it is expressed in terms of norm of the deviatoric stress $\norm{s}$ as $ f_{p\,n}\, =\,\norm{\textbf{s}_{p\,n}} \, -\, \sqrt{\frac{2}{3}}\sigma_{Y}$.

  The accumulated plastic strain $\bar{\varepsilon}^{p}_{p\,n}$ can be expressed in terms of the evolution equation of the hardening variable as
  \begin{equation}
  \dot{\bar{\varepsilon} }^{p}_{p\,n}\, =\, \sqrt{\frac{2}{3}}\norm{\dot{\varepsilon}^{p}_{p\,n}}\, =\, \dot{\gamma }_{p\,n}
  \end{equation}
 where $\dot{\gamma}_{p\,n}$  is the rate of the plastic variable.\medskip
 
The evolution equation for the plastic strain in case of isotropic associated plasticity can be expressed in terms of Lie derivative of the elastic left Cauchy Green tensor
\begin{equation}
 \mathcal{L}_{v}\textbf{b}^{e}_{p\,n}\, =\, -2\textbf{d}^{p}_{p\,n}\textbf{b}^{e}_{p\,n}\, =\, -2\dot{\gamma}_{p\,n} \frac{\partial f_{p\,n}}{\partial \boldsymbol{\tau}_{p\,n}}\textbf{b}^{e}_{p\,n}
\label{eq:evolution_equation}
 \end{equation}
 where $\textbf{d}^{p}_{p\,n}$ is the plastic rate of deformation tensor and the plastic isotropy is  modeled as $\textbf{W}^{p}\, =\, 0$ with $\textbf{W}^{p}$ as the skew symmetric part of the plastic velocity gradient. \medskip

 To model large plastic deformations, the von -Mises yield criteria  is applied alongwith linear isotropic hardening behavior (hardening modulus $H$)
 \begin{equation}
 f_{p\,n}\, =\, \norm{2\mu\boldsymbol{\varepsilon}^{e\,tr}_{p\,n}}\,-\,2\mu \Delta \gamma_{p\,n}\,-\,\sqrt{\frac{2}{3}}\left [ \sigma_{Y_{0}}\, +\, H\left ( \bar{\varepsilon}_{p\,n-1}\, +\, \sqrt{\frac{2}{3}}\Delta \gamma_{p\,n}  \right ) \right ]\leq 0
 \label{eq:von_mises_yield_2}
 \end{equation}
 where $\sigma_{Y_{0}}$ corresponds to yield stress and $\bar{\varepsilon}_{p\,n-1}$ corresponds to isotropic hardening variable computed at previous computation step. For $f\,<\,0$, the Kirchoff stresses lie in the elastic domain. But, when $f\,>\,0$ for $\Delta \gamma_{p\,n}\,=\,0$ , the yield criteria is violated and the plastic increment has to fulfill the constraint $f\,=\,0$ for Kirchoff stresses to lie on the yield surface. This can be corrected using Equation  \eqref{eq:von_mises_yield_2}. Through back transformation using the rotational tensors  as in Equation  \eqref{eq:hencky_strain_and_rotation_tensors}, the Cauchy stress tensor in Equation  \eqref{eq:algebraic_form_eq_of_motion} can be written as
 \begin{equation}
 \boldsymbol{\sigma }_{p\,n}\, =\,\textbf{Q}_{p\,n}J   \left [K \textup{tr}\, \boldsymbol{\varepsilon}^{e}_{p\,n}\mathbf{1}\, +\, 2\mu \left (\boldsymbol{\varepsilon}^{e}_{p\,n}\, -\, \textup{tr} \boldsymbol{\varepsilon}^{e}_{p\,n}\textbf{1} \right )  \right ] \textbf{Q}_{p\,n}^{\textup{T}}
 \end{equation}
 
 \subsubsection{Contact Formulation}
 
 Additionally, a contact algorithm is needed to model the copper rod striking a rigid wall. A simple contact algorithm is used assuming that the  wall is rigid and the tangential movement is frictionless. The normal gap $g_{I_{n+1}}$ of each node at the next time step can be computed as 
 \begin{equation}
 g_{I\,n+1}\, =\, \left ( \textbf{x}_{I\,n+1} \, -\, \bar{\textbf{x}}\right )\cdot \textbf{n}
 \end{equation}
 where $\bar{\textbf{x}}$ is the coordinate of the rigid plane and $\textbf{n}$ is the normal vector on that rigid plane. To enforce the non-penetration condition, a Dirichlet boundary condition is applied on the corresponding node with prescribed displacements at the next time step
 \begin{equation}
 \textbf{u}_{I\,n+1}\, =\, \textbf{x}_{I\,n}\, -\,  g_{I\,n+1}\textbf{n} 
 \end{equation}
 The above condition is only  applied when the non-penetration condition is violated $ g_{I\,n+1}\,<\,0$.  More details about formulations on two contacting deformable bodies can be found in   \citep{wriggers_contact}.

 \subsubsection{Numerical Evaluation}
 
The material parameters are chosen as $\nu\,=\,0.35$ for the Poisson ratio, $E\,=\,117.10^{9} N/m^{2}$ for the Young's modulus, $\rho_{0}\,=\,8.93\,.\,10^{3} kg/m^{3}$ for the density, $H\,=\,100.10^{6}N/m^{2}$ for the hardening modulus and $Y_{0}\,=\,400.10^{6}N/m^{2}$ for the initial yield stress. For a stable explicit time integration scheme, a computation step size of $\Delta t\,=\,4. 10^{-9}$~s is selected.

\begin{figure}[htb]	
	\centering
	{\includegraphics[width=.90\textwidth]{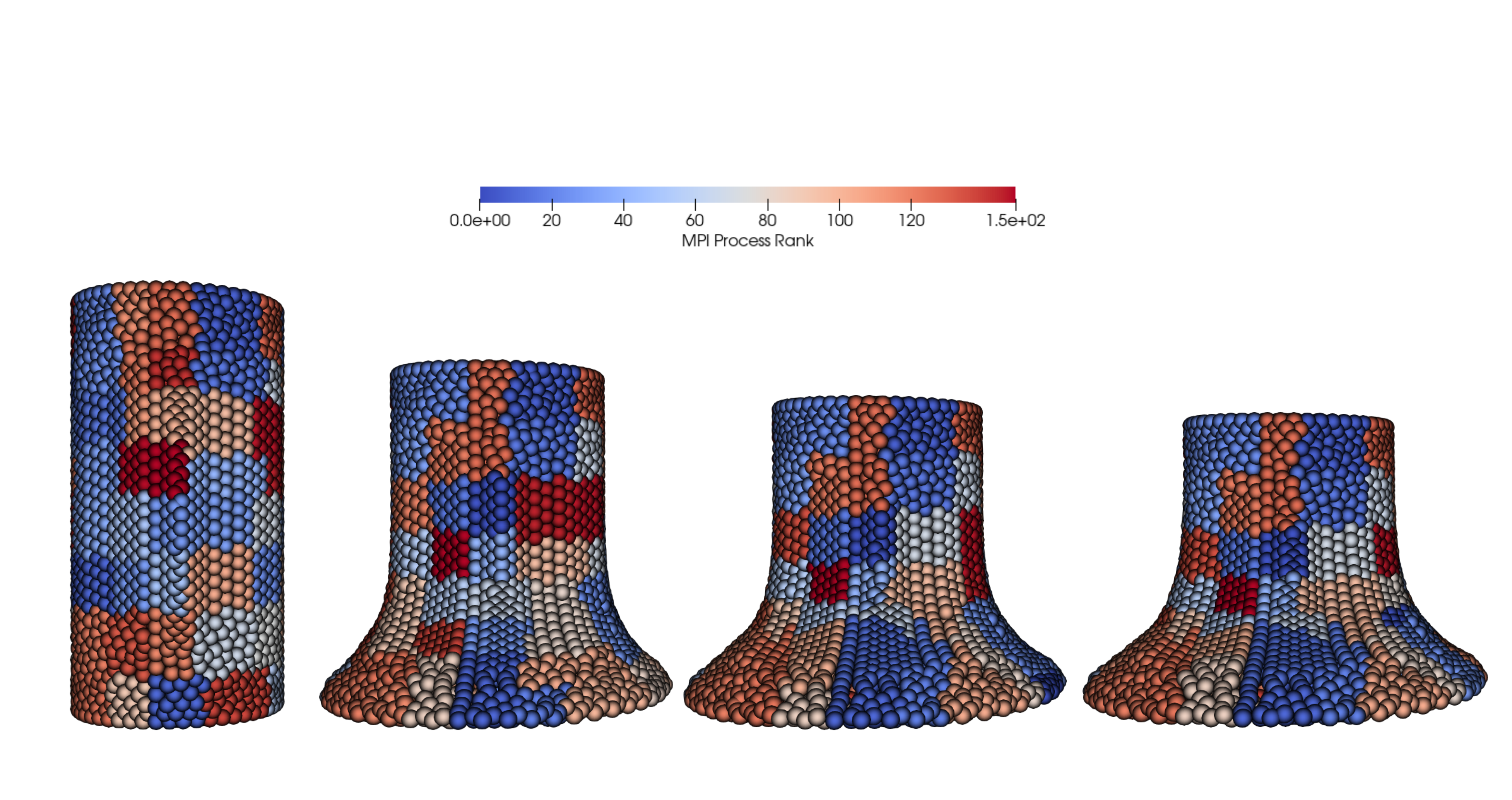}
	}\\
	\caption{Snapshots of Taylor Rod deformation alongwith nodal distribution in different sub-domains }
	\label{fig:taylor_deformation}
\end{figure}

The initial domain is set up by triangulation with the material points located at the barycenters of the tetrahedral elements. Subsequently, the initial mesh is jettisoned and the computations proceed in a meshfree manner. The model contains 5,966 nodes and  28,423 material points. The domain decomposition is performed by distributing nodes and material points across all processes with the help of Zoltan library, see Section  \ref{subsec:domain_decomposition}. Fig~\ref{fig:taylor_deformation} shows a sequence of snapshots of the Taylor rod impacting axially against a rigid boundary. Here, MPI Process Rank refers to the rank in order to identify a process, which is an integer in the range $[0,\,N-1]$ where $N$ is total number of MPI processes. \PG{Table~\ref{table:processes_vs_subdomain_size_taylor_rod} presents the average size of the subdomains in terms of own particles and halo region.}

\begin{table}[htb]
	\centering
	\begin{tabular}{|p{2.7cm}|p{2.cm}|p{2.cm}|p{2.5cm}|p{2.5cm}|}
		\hline
		\# MPI processes & \# nodes & \# MP & Av.\# halo nodes & Av.\# halo MP \\ \hline
		1                          & 5966             & 28,423     &  0   & 0        \\ \hline
		5                          & 1190             & 5683     &  434   & 2233        \\ \hline
		50                          & 120             & 569     &  209   & 1100        \\ \hline
		100                         & 60              & 285      &  145   & 772           \\ \hline
		150                         & 40              & 190       &  116    & 620          \\ \hline
		199                         & 30              & 143        &  100    & 544         \\ \hline
		239                         & 25              & 119       &  92   & 496        \\ \hline
	\end{tabular}
	\caption{Taylor rod impact: subdomain and halo average sizes (MP = material point) depending on the decomposition (halo is for the initial time step).}
	\label{table:processes_vs_subdomain_size_taylor_rod}
\end{table}



\begin{figure}[htb]
	\centering
	{\includegraphics[width=.35\textheight]{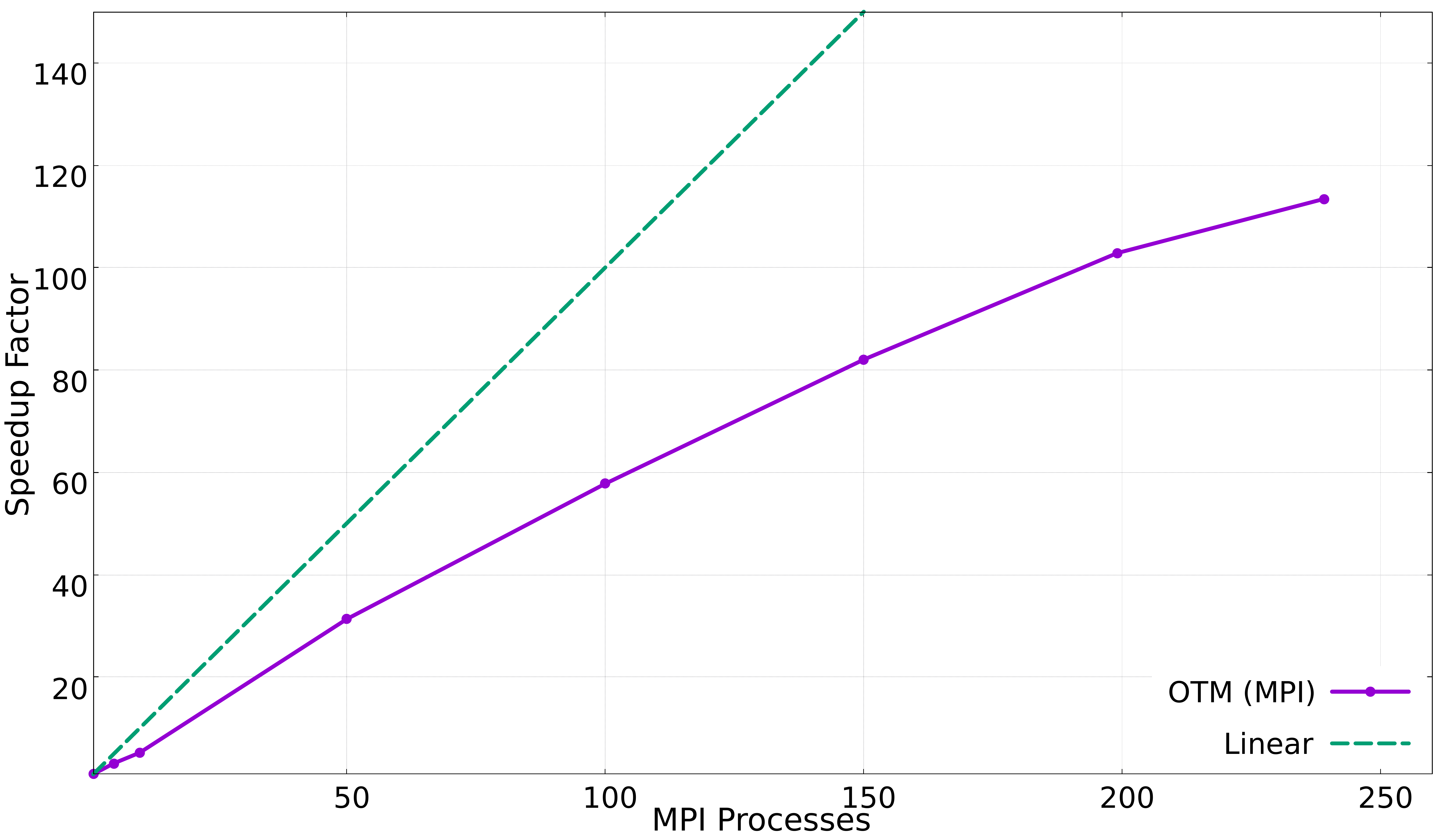}
	}\\
	\caption{Parallel performance analysis: Strong scaling for Taylor rod impact. }
	\label{fig:taylor_rod_speedup}
\end{figure}

The strong scaling studies are performed for up to 239 processes. \PG{For these studies, the code is run on the same nodes but the allocation of the cores may vary. In practice, the standard deviation for the computational time for the 3 runs is less than 5\% of the average.}  \PG{The} Parallel Performance Analysis (Fig.~\ref{fig:taylor_rod_speedup} and Table~\ref{table:parallel_performance_taylor_rod}) \PG{shows that the efficiency is about 55\% up to 150 process then it slowly decreases.} 

\begin{figure}[htb]	
	\centering
	{\includegraphics[width=.40\textheight]{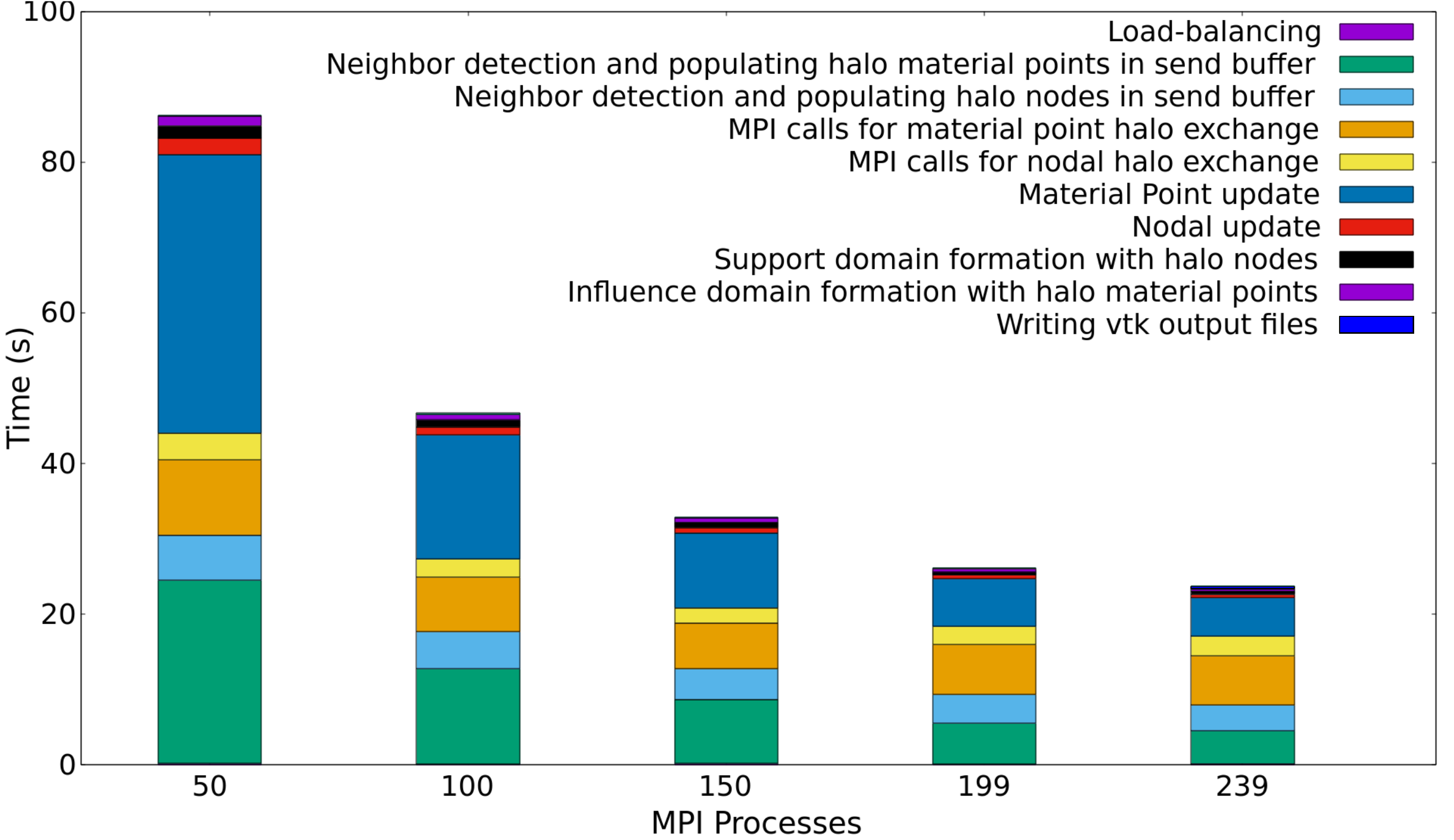}
	}\\
	\caption{Taylor rod impact: Comparisons of growth in computational and communication overhead time in strong scaling tests. }
	\label{fig:comparison_comm_and_comput_overhead}
\end{figure}

\PG{Fig~\ref{fig:comparison_comm_and_comput_overhead} tries to analyze the situation more precisely. In fact it appears that the particles computations are relatively fast, making the overtime due to data distribution significant. Then, in a first regime (number of processes larger than 2 and less than 150), adding subdomains decreases the amount of particles to treat per subdomain as well as the size of the halo particles to be exchanged, see Table~\ref{table:processes_vs_subdomain_size_taylor_rod}, making the method scale well when the reference is a parallel computation with few processes. For large number of processes, the exchange time and other incompressible stages (e.g. identification of halo particles) are dominating and the performance tends to deteriorate.}
A perspective of this work is to make a better implementation for the prediction of halo particles which are to be sent to the neighbors.


\begin{table}[htb]
	\centering
	\begin{tabular}{|l|l|l|l|}
		\hline
		Number of MPI processes & Wallclock time (s) & Speedup & Efficiency (\%) \\ \hline
		1  & 2697.26 & 1      & 100    \\ \hline
		5  & 816.67 & 3.13      & 62.6    \\ \hline
		10  & 517.82 & 5.208      & 52.08    \\ \hline
		50  & 86.19  & 31.29  & 62.58 \\ \hline
		100 & 46.69  & 57.76 & 57.76 \\ \hline
		150 & 32.88  & 82.03 & 54.68 \\ \hline
		199 & 26.23  & 102.83 & 51.97 \\ \hline
		239 & 23.78  & 113.42 & 47.45 \\ \hline
	\end{tabular}
	\caption{Performance of the parallel implementation of the OTM \PG{method} for the simulation of Taylor rod impact test.}
	\label{table:parallel_performance_taylor_rod}
\end{table}

\subsection{Application to Serrated Chip Formation Process}\label{subsec:chip_formation_process}

In the second test case, numerical modeling of chip formation is discussed. Beside physical mechanisms such as plastic deformations additionally adiabatic shear band formation and ductile fracture are involved. Only basic equations are introduced and a more detailed explanation can be found in   \citep{huang_et_al_2019}.

\subsubsection{Material Model}
The plastic deformation and the ductile fracture of the workpiece are described by the  Johnson-Cook flow stress model and Johnson-Cook fracture model respectively.
The evolution equations have the forms as shown in Equation \eqref{eq:evolution_equation}. Euler backward time integration scheme is used to solve the evolution equations based on the elastic predictor corrector return mapping algorithm, for more details, see   \citep{neto_peric_owen}.

Using von Mises plasticity, the yield function is expressed as
\begin{equation}
f^{p\, n,\, flow}\left ( \boldsymbol{\tau}_{p\,n} \right )\, =\, \sqrt{\frac{3}{2}}\left \| dev(\boldsymbol{\tau}_{p\,n}) \right \|\, -\, \sigma_{Y} \left ( \varepsilon^{p\,n}_{eq},\,\dot{\varepsilon}^{p\,n}_{eq},\,T \right )
\end{equation}
where, $\boldsymbol{\tau}_{p\,n}$ is the Kirchoff stress, $\sigma_{Y}$ the flow stress which is assumed to be a function of equivalent plastic strain rate $\dot{\varepsilon}^{p\,n}_{eq}$, equivalent plastic strain $\varepsilon^{p\,n}_{eq}$ and the temperature $T$.

Temperature increase occurs due to adiabatic heating from plastic deformation and the temperature evolution can be formulated as
\begin{equation}
\dot{T}\,=\, \beta \frac{\sigma_{v}\dot{\gamma}}{\rho\,C_{p}},\: \sigma_{v}\, =\, \sqrt{\frac{3}{2}}\left \| dev(\boldsymbol{\sigma} ) \right \|
\end{equation}
where, $\sigma_{v}$ is vonMises equivalent stress, $C_{p}$ is the heat capacity and $\beta$ is the Taylor-Quinney coefficient.

Multiplicative decomposed power form of the flow stress has been applied to consider the effects of strain hardening, strain rate hardening and thermal softening. The Johnson-Cook hardening law \citep{johnson_cook_1983} is used to capture these effects
\begin{equation}
\sigma_{Y}\, =\, \left [ A\, +\, B\left ( \varepsilon^{p\,n}_{eq}  \right )^{n} \right ] \left [ 1\, +\, Cln\left ( \frac{ \dot{\varepsilon}^{p\,n}_{eq}}{ \dot{\varepsilon}^{p\,n}_{e0}} \right ) \right ]\left [ 1\, -\, \left ( \frac{T\, -\, T_{r}}{T_{m}\,- \,T_{r} } \right )^{m} \right ]
\label{eq:johnson_cook_hardening_law}
\end{equation}
where $A$ defines the initial yield stress, $\dot{\varepsilon}^{p}_{e0}$ is the reference plain strain rate, $T_{m}$ is the melting temperature, $T_{r}$ is the room temperature and, $B$, $C$, $m$ and $n$ are additional material parameters. 

Johnson-Cook fracture model describes the separation of the chip from the workpiece and the serrated morphology on the chip upper surface. At the vicinity of the tool tip, high compression in the material and high concentration of strain occurs. Ductile fracture leads to separation of the material from the workpiece at the vicinity of the tooltip. Also,  ductile fracture at the chip upper surface can lead to the formation of serrated chips. 
Johnson-Cook fracture model is used to model the ductile fracture and to  predicts the fracture locations. When the accumulated equivalent plastic strain, $\varepsilon^{p\,n}_{eq}$ reaches the critical value,$\varepsilon^{p\,n}_{eq\, f}$, ductile fracture occurs
\begin{equation}
\varepsilon^{p\,n}_{eq}\geq\varepsilon^{p\,n}_{eq\, f} \, =\, \left [ d_{1}\, +\, d_{2}Exp\left ( d_{3}\eta   \right ) \right ] \left [ 1\, +\, d_{4}ln\left ( \frac{ \dot{\varepsilon}^{p\,n}_{eq}}{ \dot{\varepsilon}^{p\,n}_{e0}} \right ) \right ]\left [ 1\, +\,  d_{5}\frac{T\, -\, T_{r}}{T_{m}\,- \,T_{r} } \right ]
\label{eq:fracture_criteria}
\end{equation}
where $d_{1}$, $d_{2}$, $d_{3}$, $d_{4}$ and $d_{5}$ are the material parameters, $\eta$ is the stress triaxiality which is defined as
\begin{equation}
\eta \, =\, \frac{p}{\sigma_{v}}, \qquad p\, =\, \lambda tr\left ( \varepsilon ^{e} \right )
\end{equation}
where $p$ is the hydrostatic pressure and $\sigma_{v}$ is the von Mises stress.

The deformations in the chip and the workpiece during metal cutting  are driven by the cutting tool directly which moves in horizontal direction with a specific cutting depth and cutting speed. The non-penetration condition  is defined by a projection of the slave node positions from the workpiece onto the cutting tool surface
\begin{equation}
g^{N}\, =\, \left ( \textbf{x}^s\, -\, \textbf{x}^{m} \right )\cdot \textbf{n}^m \geq 0
\end{equation}
The abbreviations $g^{N}$  the normal gap,  $\textbf{x}^s$  the slave node from the workpiece $ \textbf{x}^{m}$ and $\textbf{n}^m$ are the orthogonal projection of $\textbf{x}^s$ on the tool surface and $\textbf{n}^m$ is the normal vector associated to the tool body. \medskip

The normal contact force and the stick tangential contact force can be determined by using the penalty method as
\begin{equation}
\textbf{t}^N\, =\, c_{N}\textbf{g}_{N}, \qquad \textbf{t}^T\, =\, c_{T}\textbf{g}_{T}
\end{equation}
where $c_{N}$ and $c_{T}$ are the penalty parameters. \medskip

The tangential contact force in the slip state is determined from the Coulomb friction law as
\begin{equation}
\textbf{t}^T\, =\, -\mu \norm{\textbf{t}_N}\frac{\dot{\textbf{g}}_T}{\norm{\dot{\textbf{g}}_T}}
\end{equation}
where $\mu$ is the frictional coefficient. Further details can be found in   \citep{huang_et_al_2019}.

\subsubsection{Numerical Evaluation}

\begin{figure}[htb]	
	\centering
	{\includegraphics[width=.50\textwidth]{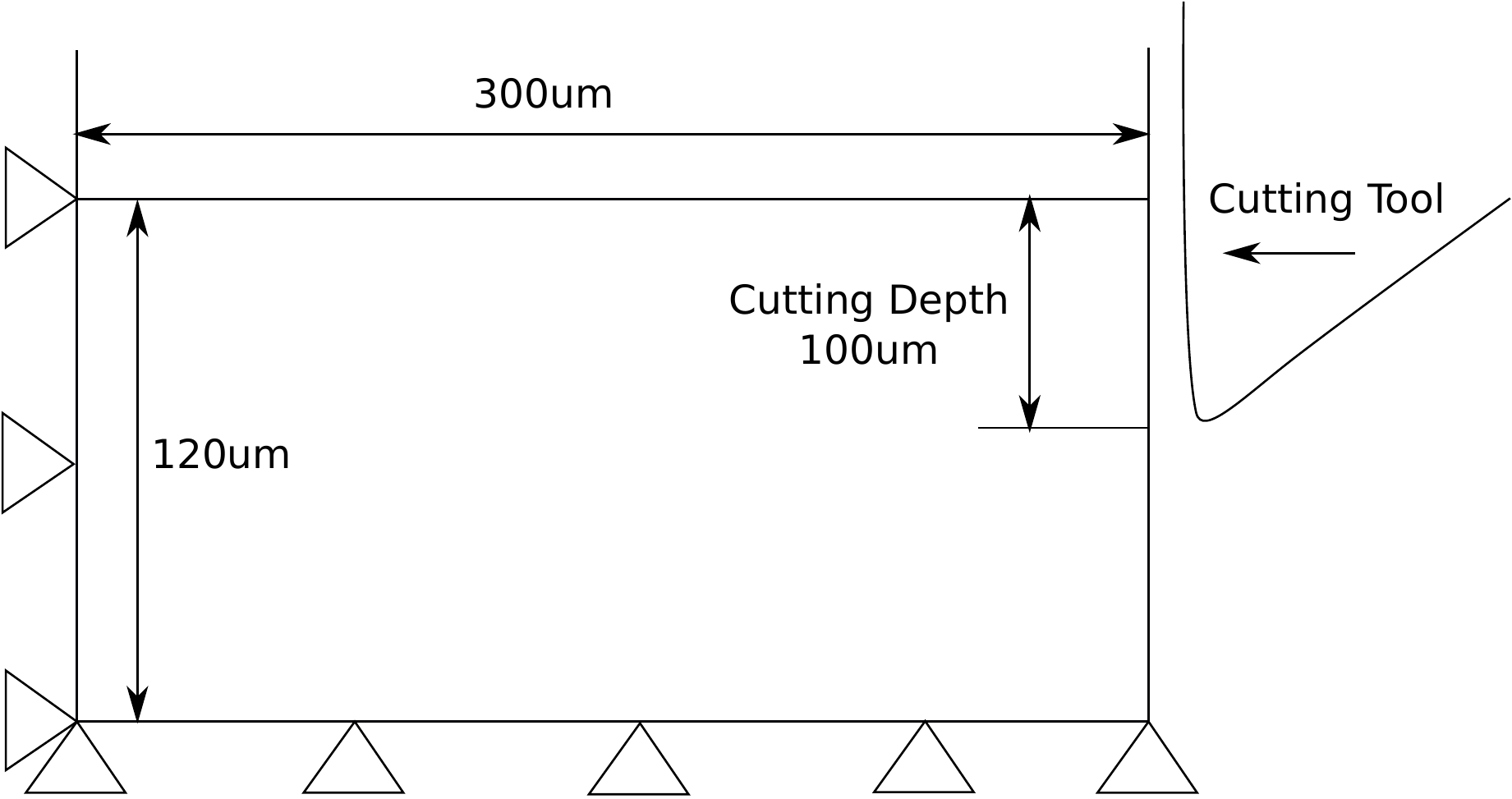}
	}\\
	\caption{Geometrical model for metal cutting.}
	\label{fig:machining_geometrical_model}
\end{figure}

Ti6Al4V alloy is used as the workpiece material. The material parameters of the constitutive equations~(\ref{eq:johnson_cook_hardening_law}) and (\ref{eq:fracture_criteria})  can be found in    \citep{huang_et_al_2019}. The workpiece has a length and height of 300 $\mu$m and 120 $\mu$m respectively, see Fig \ref{fig:machining_geometrical_model}. The cutting depth is 100 $\mu$m. The cutting tool is treated as  rigid body with tool radius of 2 $\mu$m and rake angle of 0$^{\circ}$. For the tool-chip contact modeling, the friction coefficient is set as 0.8. For the workpiece, the melting temperature $T_{m}$ and the initial temperature $T_{r}$ is set as 1630$^{\circ}$C and 25$^{\circ}$C respectively. For a stable explicit time integration scheme, a time step size of $\Delta t\,=\,10^{-10}$~s is selected.

\begin{figure}[htb]	
	\centering
	{\includegraphics[width=.90\textwidth]{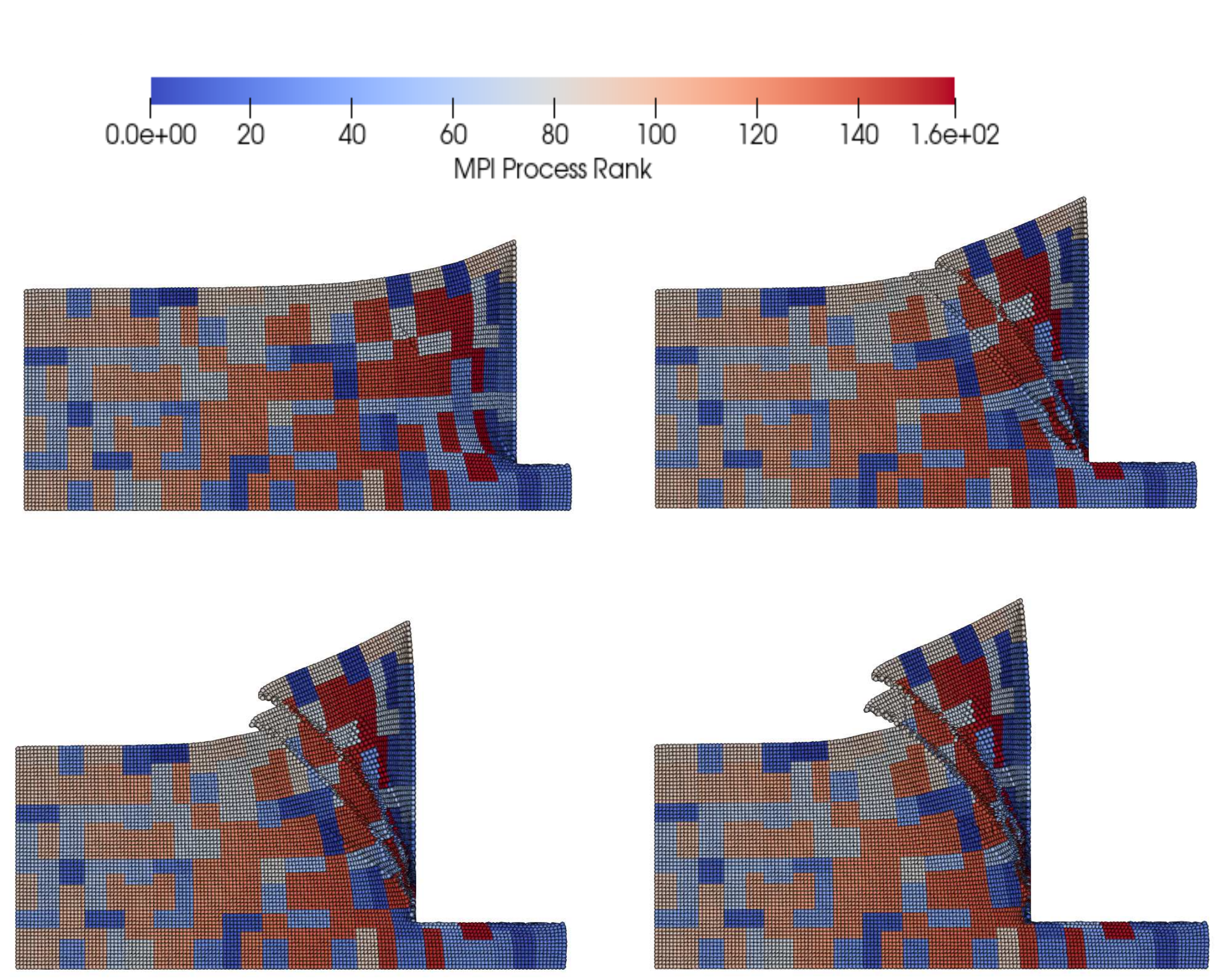}
	}\\
	\caption{Snapshots of serrated chip formation process alongwith nodal distribution in different sub-domains. }
	\label{fig:machining_deformation_steps}
\end{figure}

\begin{figure}[htb]	
	\centering
	{\includegraphics[width=.35\textheight]{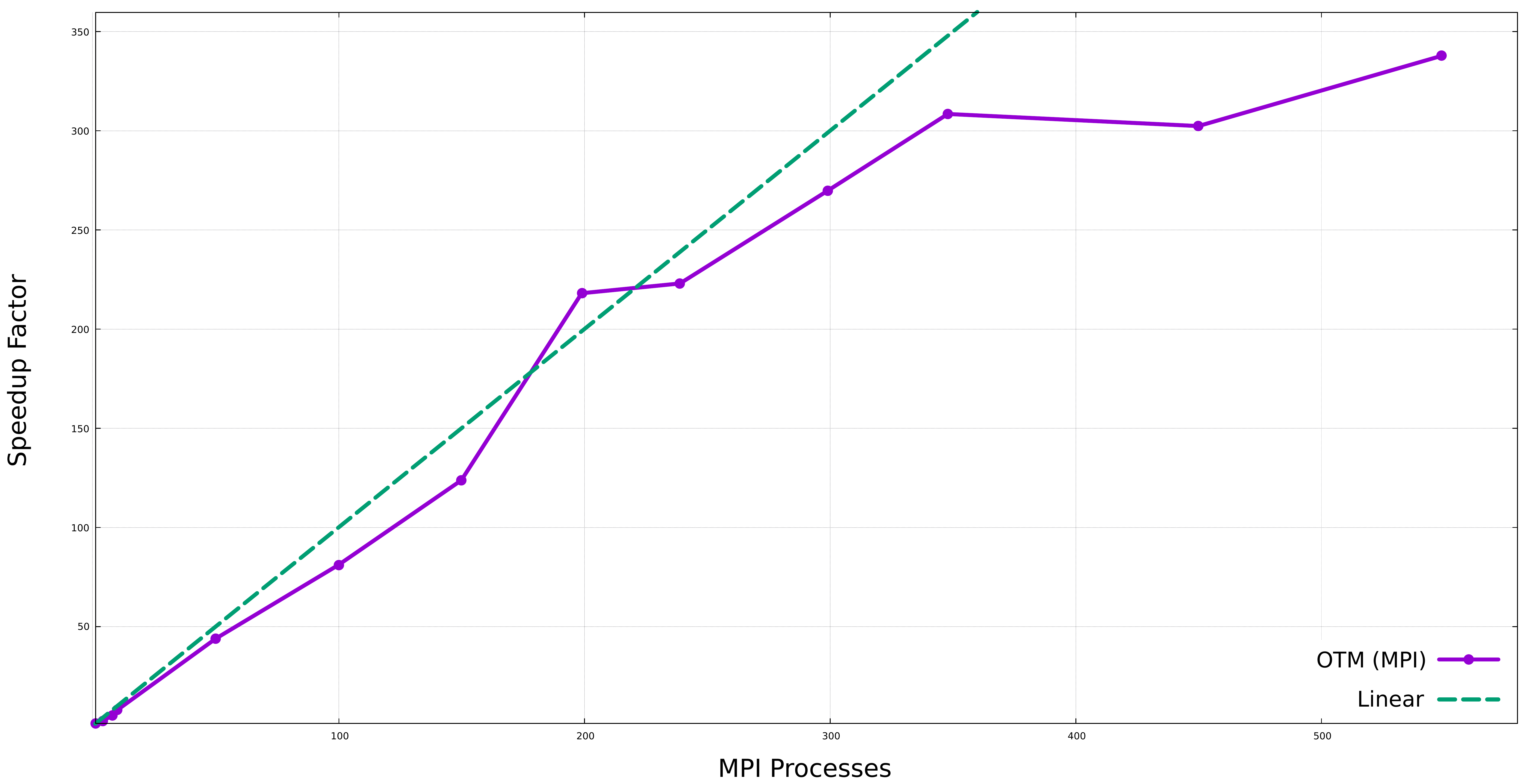}
	}
	\caption{Parallel performance analysis: Strong scaling for serrated chip formation process. }
	\label{fig:machining_speedup_1}
\end{figure}

\begin{figure}[htb]	
	\centering
	{\includegraphics[width=.40\textheight]{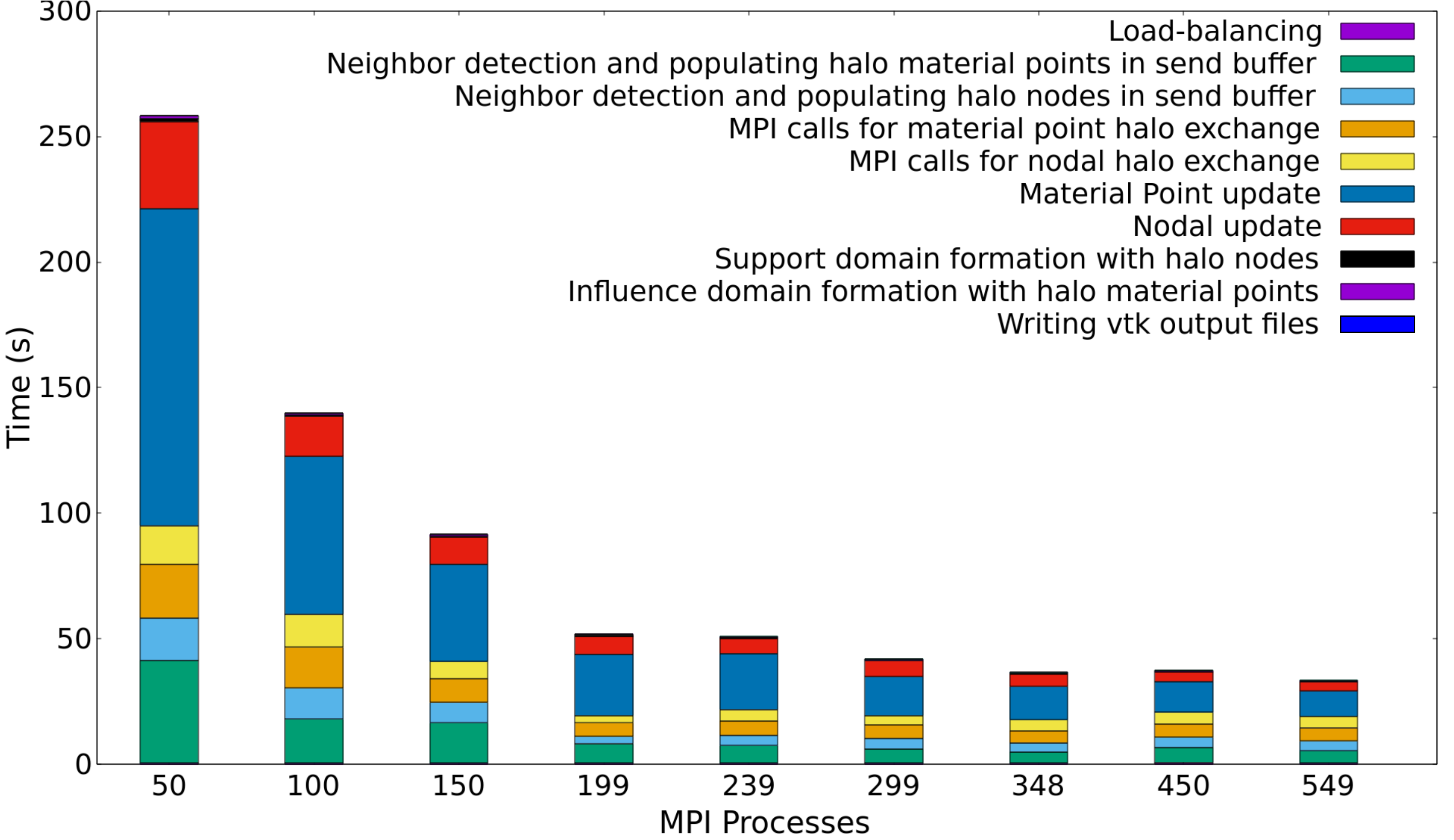}
	}
	\caption{Serrated chip formation process: Comparisons of growth in computational and communication overhead time in strong scaling tests. }
	\label{fig:comparison_overheads_machining}
\end{figure}

The model consists of 27,417 nodes and 107,975 material points. Their distribution in subdomains is presented in Table~\ref{table:processes_vs_subdomain_size_machining}.
In this example, the scalability performance of the multiprocessing approach in the numerical solutions of large deformation problem is investigated. In Fig  \ref{fig:machining_deformation_steps}, the sequence of the serrated chip formation process is shown together with  the corresponding nodal distribution across the sub-domains. {Strong scaling studies are conducted for up to 549 processes, see Figure~\ref{fig:machining_speedup_1} and Table~\ref{table:parallel_performance_machining_1}, as well as Figure~\ref{fig:comparison_overheads_machining} for a more detailed analysis.}

We observe that the parallel efficiency is better than from previous test case. The speedup is excellent up to 400 processes. This difference can be explained by the fact that the problem is mostly 2D  and each subdomain possesses at most 8 neighbors, which limits the communications. Note that the 199-process case exceeds linear behavior, but this remains in range of measurement variability.

\begin{table}[htb]
	\centering
	\begin{tabular}{|p{3cm}|p{2cm}|p{2cm}|p{2.5cm}|p{2.5cm}|}
		\hline
		\# MPI processes & \# nodes & \# MP & Av.\# halo nodes & Av.\# halo MP \\ \hline
		1                          & 27,417             & 107,975      & 0&     0           \\ \hline
		4                          & 6852             & 26,993      & 288&     1047           \\ \hline
		8                          & 3428             & 13,497      & 214&     864           \\ \hline
		50                          & 549             & 2160      & 98&     417           \\ \hline
		100                         & 275             & 1080      & 76&      293          \\ \hline
		150                         & 183             & 720       & 63&      250          \\ \hline
		199                         & 138             & 543       & 57 &     225           \\ \hline
		239                         & 115             & 452       & 51&      204          \\ \hline
		299                         & 92              & 362       & 47&      185          \\ \hline
		348                         & 79              & 311       & 43&      170          \\ \hline
		450                         & 61              & 240        & 39&     151          \\ \hline
		549                         & 50              & 197        & 36&     137          \\ \hline
	\end{tabular}
	\caption{Serrated chip formation process :subdomain and halo average sizes (MP = material point) depending on the decomposition (halo is for the initial time step).}
	\label{table:processes_vs_subdomain_size_machining}
\end{table}

\begin{table}[htb]
	\centering
	\begin{tabular}{|l|l|l|l|}
		\hline
		Number of MPI processes & Wallclock time (s) & Speedup & Efficiency (\%) \\ \hline
		1   & 11348.8932 & 1.0     & 100   \\ \hline
		4   & 4514.898 & 2.51    & 62.84   \\ \hline
		8   & 2224.957 & 5.1     & 63.75   \\ \hline
		10  & 1432.27 & 7.923      & 79.23    \\ \hline
		50  & 258.27  & 43.94  & 87.88 \\ \hline
		100 & 139.88  & 81.13  & 81.13  \\ \hline
		150 & 91.66   & 123.81 & 82.54 \\ \hline
		199 & 52.017   & 218.17 & 109.63 \\ \hline
		239 & 50.877   & 223.06 & 93.33 \\ \hline
		299 & 42.063   & 269.80 & 90.23 \\ \hline
		348 & 36.780   & 308.56 & 88.66 \\ \hline
		450 & 37.520   & 302.47 & 67.21 \\ \hline
		549 & 33.589   & 337.87 & 61.54 \\ \hline
	\end{tabular}
	\caption{Performance of the parallel implementation of the OTM \PG{method} for the simulation of the serrated chip formation process.}
	\label{table:parallel_performance_machining_1}
\end{table}

\section{Conclusions}
\label{sec:conclusions}

An  OTM algorithm for large deformations, parallelized using MPI with an objective for scalability on large scale CPU clusters has been presented. The consistency and robustness of this algorithm is demonstrated by two examples showing large deformation. Strong scaling studies were conducted. 
Implementation of dynamic halo regions have shown to improve the scalability by its ability to handle variable workloads and eliminating the storage issues related to fixed-size arrays.  With the increase in number of processes, good scalability is observed \PG{for the 2D Serrated Chip Formation Process example}. The communication costs decreases significantly and asymptotically even though more subdomain interfaces are present leading to increase in number of halo particles. The second advantage is the efficient data management strategy using advanced STL container adapted to fulfill various functionalities of data structure modifications. Flexible handling of data structures for two types of particles (nodes and material points) resulted in reduction of computational costs. Together with localized computation within each sub-domain by using nearest neighborhood collectives for both nodes and material points this approach leads to  scalable results. \PG{ Since the method is explicit in time, computations are very simple and with low granularity. We may study new ordering of operations (starting from the innermost particles) so that computations could be started while the halo are being exchanged, in the spirit of \citep{cornelis_et_al_2018}. We could also consider larger halo with several time steps being computed without synchronizations, or even fully asynchronous computations, see for example \citep{magoules_2018}. Anyhow, the first improvement to be implemented is a mixed MPI/OpenMP approach to parallelism.}

\section{Acknowledgements}
\label{sec:acknowledgements}
Funding supports from Deutsche Forschungsgemeinschaft DFG within the research training centre ViVaCE (IRTG 1627), French-German doctoral college 'Sophisticated Numerical and Testing Approaches' (SNTA) and Graduierten Akademie-Leibniz Universität Hannover is gratefully acknowledged.


\small

\bibliographystyle{plainnat}
\bibliography{paper}

\end{document}